\documentclass{ws-rv975x65}
\usepackage{ws-rv-van}             
\makeindex

\newcommand{\be}{\begin{equation}}
\newcommand{\ee}{\end{equation}}
\newcommand{\bea}{\begin{eqnarray}}
\newcommand{\eea}{\end{eqnarray}}
\newcommand{\br}{{\bf r}}

\usepackage{graphicx}
\usepackage{mathrsfs}
\usepackage{bm}
\usepackage{chapterbib}
\usepackage{amsmath}
\usepackage{amssymb}
\usepackage{amsfonts}
\usepackage{upgreek}
\usepackage{wasysym}
\usepackage{mathbbol}

\begin{document}

\chapter[Anderson Transitions] {Anderson Transitions: Criticality,
  Symmetries, and Topologies}

\author[A. D. Mirlin, F. Evers, I. V. Gornyi, and P. M. Ostrovsky]
{A. D. Mirlin\footnote{Also at Institut f\"ur Theorie der
kondensierten Materie,
 Karlsruhe Institute of Technology, 76128 Karlsruhe, Germany and
 Petersburg Nuclear Physics Institute, 
 188300 St.~Petersburg, Russia},  F. Evers, I. V. Gornyi\footnote{Also
 at A.F.~Ioffe Physico-Technical Institute, 
 194021 St.~Petersburg, Russia}, and P. M. Ostrovsky\footnote{Also at
 L.~D.~Landau Institute for Theoretical Physics RAS, 
 119334 Moscow, Russia}}

\address{Institut f\"ur Nanotechnologie,\\ Karlsruhe Institute of Technology,
 76021 Karlsruhe, Germany}

\begin{abstract}

The physics of Anderson transitions between
localized and metallic phases in disordered systems is reviewed.
We focus on the character of criticality as well as on underlying
symmetries and topologies that are crucial for understanding phase
diagrams and the critical behavior.

\end{abstract}

\body

\section{Introduction}
\label{sec1}

Quantum interference can
completely suppress the diffusion of a particle in random potential, a
phenomenon known as {\em Anderson localization}.\cite{anderson58}
For a given energy and disorder strength the quantum
states are either all localized or all delocalized.
This implies the existence of {\em Anderson transitions} between
localized and metallic phases in disordered electronic systems.
A great progress in understanding of the corresponding
physics was achieved in the seventies and the eighties, due to the
developments of
scaling theory and field-theoretical approaches to localization,
which demonstrated connections between the Anderson transition and
conventional second-order phase transitions; see review
articles\cite{lee85,kramer93,huckestein95} and the book \cite{efetov97}.

During the last 15 years considerable progress in the field has
been made in several research directions. This has strongly advanced the
understanding of the physics of Anderson localization and associated
quantum phase transitions and
allows us to view it nowadays in a considerably broader and more
general context\cite{evers08}.

First, the {\it symmetry} classification of disordered systems has been
completed. It has been understood that a complete set of random matrix
theories includes,
in addition to the three Wigner-Dyson classes, three chiral
ensembles and four Bogoliubov-de Gennes ensembles \cite{altland97}.
Zirnbauer has established a relation between
random matrix theories, $\sigma$-models
and Cartan's classification of symmetric spaces,
which provides the mathematical basis for the statement of completeness
of the classification \cite{zirnbauer96b}.
The additional ensembles are characterized by one of the additional
symmetries -- the chiral or the particle-hole one.
The field theories ($\sigma$-models)
associated with these new symmetry classes have in fact been
considered already in the eighties.
However, it was only after their physical
significance had been better understood that the new symmetry classes were
studied systematically.

Second, the classification of fixed points governing the localization
transitions in disordered metals was found to be much richer than that of
symmetries of random matrix ensembles (or field theories).
The first prominent example of this was in fact given 25 years ago by
Pruisken \cite{pruisken84} who showed that the quantum Hall transition is
described by a $\sigma$-model with an additional, topological, term. However,
it is only recently that the variety of types of criticality---and, in
particular, the impact of {\it topology}---was fully appreciated. Recent
experimental discoveries of graphene and topological insulators have
greatly boosted the research activity in this direction.

Third, an important progress in understanding the statistics of
wave functions at criticality has been made. Critical wave functions
show very strong fluctuations and long-range correlations that are
characterized by {\it multifractality}
\cite{janssen84,huckestein95,mirlin00a,evers08} implying the presence of
infinitely many relevant operators. The spectrum of multifractal
exponents constitutes a crucially important characteristics of the
Anderson transition fixed point. The understanding of general
properties of the statistics of critical wave functions and their
multifractality was complemented by a detailed study -- analytical and
numerical -- for a number of localization critical points, such as
conventional Anderson transition in various dimensionalities, 2D Dirac
fermions in a random vector potential, integer quantum Hall effect
(IQHE), spin quantum Hall effect (SQHE), 2D
symplectic-class Anderson transition, as well as the power-law
random banded matrix (PRBM) model.

Fourth, for several types of Anderson transitions,
very detailed studies using both analytical and numerical tools have been
performed. As a result,
a fairly comprehensive {\it quantitative understanding}
of the localization critical phenomena has been achieved. In
particular, the PRBM model, which can be viewed as a 1D
system with long-range hopping, has been analytically solved on its critical
line \cite{mirlin96,mirlin00a,evers08}. This allowed a detailed study of
the statistics of wave functions and energy levels at criticality.
The PRBM model serves at present
as a ``toy model'' for the Anderson criticality.
This model possesses a truly marginal coupling, thus yielding a line of
critical points and allowing to study the evolution of critical properties in
the whole range from weak- to strong-coupling fixed points. Further recent
advances in quantitative understanding of the critical behavior of
Anderson transitions are related to exploration of network models of
IQHE  and its ``relatives'' from other symmetry classes, development
of theories of disordered Dirac fermions, as well as large progress in
numerical simulations.

Finally, important advances have been achieved in understanding the
impact of the {\it electron-electron interaction} on Anderson transitions.
While this article mainly deals with non-interacting systems, we will
discuss most prominent manifestations of the interaction in
Sec.~\ref{s7.1} and \ref{s8.2}.

This article presents an overview of
field with an emphasis on recent developments. The main
focus is put on conceptual issues related to phase diagrams, the
nature of criticality, and the role of underlying symmetries and
topologies. For a more detailed exposition of the physics of
particular Anderson transition points and an extended bibliography
the reader is referred to a
recent review, Ref.~\refcite{evers08}.

\section{Anderson transitions in conventional symmetry classes}
\label{s2}

\subsection{Scaling theory, observables, and critical behavior}
\label{s2.1}

When the energy or the disorder strength is varied, the system can
undergo a transition from the metallic phase with delocalized
eigenstates to the insulating phase, where eigenfunctions are
exponentially localized\cite{anderson58}, \be \label{e2.1}
|\psi^2(\bf r)| \sim \exp(-|\bf r - {\bf r_0}|/\xi), \ee and $\xi$
is the localization length. The character of this transition
remained, however, unclear for roughly 20 years, until Wegner
conjectured, developing earlier ideas of
Thouless\cite{thouless74}, a close connection between the Anderson
transition and  the scaling theory of critical
phenomena.\cite{wegner76}   Three years later, Abrahams, Anderson,
Licciardello, and Ramakrishnan formulated a {\em scaling theory}
of localization\cite{abrahams79}, which describes the flow of the
dimensionless conductance $g$ with the system size $L$, \be
\label{e2.2} d\ln g /d \ln L = \beta(g). \ee This phenomenological
theory was put on a solid basis after Wegner's discovery of the
field-theoretical description of the localization problem in terms
of a nonlinear $\sigma$-model\cite{wegner79}, Sec.~\ref{s2.2}.
This paved the way for the resummation of singularities in
perturbation theory at or near two
dimensions\cite{gorkov79,vollhardt80} and allowed to cast the
scaling in the systematic form of a field-theoretical
renormalization group (RG). A microscopic derivation of the
$\sigma$-model worked out in a number of papers
\cite{schaefer80,juengling80,efetov80} has completed a case for it
as the field theory of the Anderson localization.

To analyze the transition, one starts from
the Hamiltonian $\hat{H}$  consisting of the free part
$\hat{H}_0$ and the disorder potential $U({\bf r})$:
\begin{equation}
\hat{H}=\hat{H}_0+U({\bf r})\ ;\qquad \hat{H}_0 =
\hat{{\bf p}}^2 / 2m.
\label{e2.2a}
\end{equation}
The disorder is defined by the correlation function $\langle U({\bf
  r})U({\bf r'})\rangle$; we can assume it to be of the white-noise
type for definiteness,
\begin{equation}
\langle U({\bf r})U({\bf r'})\rangle=
(2\pi\rho\tau)^{-1}\delta({\bf r}-{\bf r'}).
\label{e2.2b}
\end{equation}
Here $\rho$ is the density of states, $\tau$ the mean free
time and $\langle\ldots\rangle$ denote the disorder average.
Models with finite-range and/or
anisotropic disorder correlations
are equivalent with respect to the long-time, long-distance
behavior to the white noise model
with renormalized parameters
(tensor of diffusion coefficients)\cite{woelfle84}.


The physical observables whose scaling at the transition point is of primary
importance is the localization length $\xi$
on the insulating side (say, $E<E_c$)
and the DC conductivity $\sigma$ on the metallic side ($E>E_c$),
\be
\label{e2.3}
\xi \propto (E_c-E)^{-\nu}, \qquad \sigma \propto (E-E_c)^s.
\ee
The corresponding critical indices $\nu$ and $s$ satisfy the scaling
relation\cite{wegner76}
$s=\nu (d-2)$.

On a technical level, the transition manifests
itself in a change of the behavior of the diffusion propagator,
\be
\label{e2.4}
\Pi({\bf r_1},{\bf r_2}; \omega) = \langle G^R_{E+\omega/2}({\bf r_1}, {\bf
  r_2})  G^A_{E-\omega/2}({\bf r_2}, {\bf r_1})\rangle,
\ee where $G^R$, $G^A$ are retarded and advanced Green functions,
\be \label{e2.5} G^{R,A}_{E}({\bf r}, {\bf r'}) = \langle {\bf r}
| (E-\hat{H}\pm i\eta)^{-1}| {\bf r'}\rangle, \qquad \eta\to +0.
\ee In the delocalized regime $\Pi$ has the familiar diffusion
form (in the momentum space), \be \label{e2.6} \Pi({\bf q},\omega)
= 2\pi \rho(E) / (Dq^2-i\omega), \ee where $\rho$ is the density
of states (DOS) and $D$ is the diffusion constant, related to the
conductivity via the Einstein relation $\sigma = e^2 \rho D$. In
the insulating phase, the propagator ceases to have the Goldstone
form (\ref{e2.6}) and becomes massive, \be \label{e2.7} \Pi ({\bf
r_1},{\bf r_2}; \omega) \simeq \frac{2\pi\rho}{-i\omega} {\cal
  F} (|{\bf r_1}-{\bf r_2}|/\xi),
\ee
with the function ${\cal F}({\bf r})$ decaying exponentially on the
scale of the localization length, ${\cal F}(r/\xi) \sim \exp(-r/\xi)$.
It is worth emphasizing that the localization length $\xi$ obtained
from the averaged correlation function $\Pi = \langle G^RG^A\rangle$,
Eq.~(\ref{e2.4}), is in general different from the one governing the
exponential decay of the typical value $\Pi_{\rm typ} = \exp \langle
\ln G^RG^A\rangle$. For example, in quasi-1D systems the
two lengths differ by a factor of four \cite{mirlin00a}. However, this is
usually not important for the definition of the
critical index $\nu$.
We will return to observables related to critical
fluctuations of wave functions and discuss the corresponding family of
critical exponents in Sec.~\ref{s2.3}.

\subsection{Field-theoretical description}
\label{s2.2}

\subsubsection{Effective field theory: Non-linear $\sigma$-model}
\label{s2.2.1}

In the original derivation of the $\sigma$-model
\cite{wegner79,schaefer80,juengling80,efetov80},  the
replica trick was used to perform the disorder
averaging. Within this approach, $n$ copies of the system are considered, with
fields $\phi_\alpha$, $\alpha =1,\ldots,n$ describing the particles,
and the replica limit $n\to 0$ is taken in the end. The resulting
$\sigma$-model is defined on the $n\to 0$ limit of either
non-compact or compact symmetric space, depending on whether the
fields $\phi_\alpha$ are considered as bosonic or fermionic. As an
example, for the unitary symmetry class (A), which
corresponds to a system with broken time-reversal invariance, the
$\sigma$-model target manifold is $\text{U}(n,n)/\text{U}(n)\times
\text{U}(n)$ in the first case
and $\text{U}(2n)/\text{U}(n)\times \text{U}(n)$ in the second case,
with $n \to 0$. A
supersymmetric formulation given by Efetov\cite{efetov97} combines
fermionic and bosonic degrees of freedom, with the field
$\Phi$ becoming a supervector. The resulting $\sigma$-model is defined
on a supersymmetric coset space,
e.g. $\text{U}(1,1|2)/\text{U}(1|1)\times \text{U}(1|1)$
for the unitary class. This manifold combines compact and non-compact
features and represents a product of the hyperboloid $\text{H}^2 =
\text{U}(1,1)/\text{U}(1)\times \text{U}(1)$ and the sphere
$\text{S}^2=\text{U}(2)/\text{U}(1)\times \text{U}(1)$
``dressed'' by anticommuting (Grassmannian) variables.
While being equivalent
to the replica version on the level of the perturbation theory
(including its RG resummation), the supersymmetry formalism allows also for a
non-perturbative treatment of the theory, which is particularly
important for the analysis of the energy level and eigenfunction
statistics, properties of quasi-1D systems, topological effects,
etc.~\cite{efetov97,guhr98,mirlin00a,zirnbauer04}.

Focusing on the unitary symmetry class, the expression for the
propagator $\Pi$, Eq.~(\ref{e2.4}) is obtained as
\be
\label{e2.11}
\Pi({\bf r_1},{\bf r_2}; \omega) = \int DQ\:  Q_{12}^{bb}({\bf r_1})
Q_{21}^{bb}({\bf r_2}) e^{-S[Q]},
\ee
where $S[Q]$ is the $\sigma$-model action
\be
\label{e2.12}
S[Q] = \frac{\pi\rho}{4}\int d^d{\bf r} \:
\mbox{Str}\: [-D(\nabla Q)^2 - 2i\omega\Lambda Q].
\ee
Here $Q=T^{-1}\Lambda T$ is a $4\times 4$ supermatrix that
satisfies the condition $Q^2=1$ and belongs to the $\sigma$-model target space
described above, $\Lambda=\mbox{diag}\{1,1,-1,-1\}$, and Str denotes
the supertrace. The size $4$ of the matrix is due to
(i) two types of the Green functions (advanced and retarded),
and (ii) necessity to introduce bosonic and fermionic
degrees of freedom to represent these Green's function in terms of a
functional integral. The matrix $Q$ consists thus of four $2\times 2$
blocks according to its advanced-retarded structure,
each of them being a supermatrix in the boson-fermion space. In
particular, $Q_{12}^{bb}$ is the boson-boson element of the RA block,
and so on.
One can also consider an average of the product of $n$
retarded and $n$ advanced Green functions, which will generate a
$\sigma$-model defined on a larger manifold, with the base being a
product of $\text{U}(n,n)/\text{U}(n)\times \text{U}(n)$ and
$\text{U}(2n)/\text{U}(n)\times \text{U}(n)$
(these are the same structures as in the replica formalism, but now
{\it without} the $n\to 0$ limit).

For other symmetry classes, the symmetry of the $\sigma$-model is
different but the general picture is the same. For example, for the
orthogonal class (AI) the $8\times 8$
$Q$-matrices span the manifold whose base is
the product of the non-compact space
$\text{O}(2,2)/\text{O}(2)\times \text{O}(2)$ and the
compact space $\text{Sp}(4)/\text{Sp}(2)\times \text{Sp}(2)$.
The $\sigma$-model symmetric
spaces for all the classes (Wigner-Dyson as well as unconventional)
are listed in Sec.~\ref{s:SymDisSys}.

\subsubsection{RG in $2+\epsilon$ dimensions; $\epsilon$-expansion}
\label{s2.2.2}

The $\sigma$-model is the effective low-momentum, low-frequency theory
of the problem, describing the dynamics of interacting soft modes --
diffusons and cooperons. Its RG treatment yields a flow equation of
the form (\ref{e2.2}), thus justifying the scaling theory of
localization. The $\beta$-function $\beta(t) \equiv - dt/d\ln L$
can be calculated perturbatively in
the coupling constant $t$ inversely proportional to the dimensionless
conductance,  $t=1/2\pi g$.\footnote{For spinful systems, $g$ here
does not include summation over spin projections.}
 This allows one to get the
$\epsilon$-expansion for the critical exponents in $2+\epsilon$
dimensions, where the transition takes place at $t_*\ll 1$. In
particular, for the orthogonal symmetry class (AI) one finds
\cite{wegner88}
\be
\label{e2.13}
\beta(t) = \epsilon t - 2t^2 - 12 \zeta(3)t^5 +O(t^6).
\ee
The transition point $t_*$ is given by the zero of the $\beta(t)$,
\be
\label{e2.14}
t_* = {\epsilon/2}  - (3/8) \: \zeta(3) \epsilon^4 + O(\epsilon^5).
\ee
The localization length exponent $\nu$ is determined by the derivative
\be
\label{e2.15}
\nu = -1/\beta'(t_*) = \epsilon^{-1} - (9/ 4)\: \zeta(3)\epsilon^2 +
O(\epsilon^3),
\ee
 and the conductivity exponent $s$ is
\be
\label{e2.16}
s = \nu \epsilon = 1 - (9/4)\: \zeta(3)\epsilon^3 +
O(\epsilon^4).
\ee
Numerical simulations of localization on fractals with dimensionality
slightly above 2 give the behavior of $\nu$ that is in good agreement with
Eq.~(\ref{e2.15}) \cite{schreiber96}.
For the unitary symmetry class (A), the corresponding results read
\bea
\label{e2.17}
&& \hspace*{-1cm} \beta(t) = \epsilon t - 2t^3 - 6t^5 +O(t^7); \\
&&  \hspace*{-1cm}  t_* = \left(\epsilon/2\right)^{1/2} - (3/2)
 \left(\epsilon/ 2\right)^{3/2} + O(\epsilon^{5/2});
\label{e2.18} \\
&& \hspace*{-1cm}
  \nu = 1/2\epsilon - 3/ 4 +O(\epsilon)\ ; \qquad
s = 1/2 - (3/4)\epsilon +O(\epsilon^2).
\label{e2.19}
\eea
In 2D ($\epsilon=0$) the fixed point $t_*$ in both cases
becomes zero: $\beta(t)$
is negative for any $t>0$, implying that all states are localized. The
situation is qualitatively different for the third---symplectic---Wigner-Dyson class. 
The corresponding $\beta$-function is related to that for the
orthogonal class via $\beta_{\rm Sp}(t) = - 2 \beta_{\rm O} (-t/2)$,
yielding\footnote{Here $t=1/\pi g$, where $g$ is the total
  conductance of the spinful system.}
\be
\label{e2.20}
\beta(t) = \epsilon t + t^2 - (3/4)\: \zeta(3) t^5 +O(t^6).
\ee
In 2D  the $\beta$-function (\ref{e2.20}) is positive at
sufficiently small $t$,
implying the existence of a truly metallic phase at $t< t_*$, with an Anderson
transition at certain $t_* \sim 1$. This peculiarity
of the symplectic class
represents one of mechanisms of the emergence of criticality in 2D,
see Sec.~\ref{s6.1}.
The $\beta$-functions of unconventional
symmetry classes will  be discussed in Sec.~\ref{s4.6}.

\subsection{Critical wave functions: Multifractality}
\label{s2.3}

\subsubsection{Scaling of inverse participation ratios and
  correlations at criticality}
\label{s2.3.1}

Multifractality of wave functions, describing their strong
fluctuations at criticality, is a striking feature of the Anderson
transitions \cite{wegner80,castellani86}.
Multifractality as a concept has been introduced by
Mandelbrot\cite{mandelbrot74}. Multifractal structures
are characterized by an infinite set of
critical exponents describing the scaling of the moments of some
distribution. This feature has been observed in various complex
objects, such as the energy dissipating set in turbulence, strange
attractors in chaotic dynamical systems, and the growth probability
distribution in diffusion-limited aggregation. For the present
problem, the underlying normalized measure is just $|\psi^2({\bf r})|$ and
the corresponding moments are the inverse participation ratios (IPR)
\footnote{Strictly speaking, $P_q$ as defined by Eq.~(\ref{e2.22}),
diverges for sufficiently negative $q$ ($q\le -1/2$ for real $\psi$
and $q\le -3/2$ for complex $\psi$), because of zeros of wave
functions related to their oscillations on the scale of the wave
length. To find $\tau_q$ for such negative $q$, one should first
smooth $|\psi^2|$ by averaging over some microscopic volume (block  of
several neighboring sites in the discrete version).}
\be
\label{e2.22}
P_q = \int d^d{\bf r} |\psi({\bf r})|^{2q}.
\ee
At criticality, $P_q$ show an anomalous scaling with the system size
$L$,
\be
\label{e2.23}
\langle P_q\rangle = L^d  \langle|\psi({\bf r})|^{2q}\rangle \sim
L^{-\tau_q},
\ee
governed by a continuous set of exponents $\tau_q$. One often
introduces fractal dimensions $D_q$ via $\tau_q{=}D_q(q-1)$. In a metal
$D_q{=}d$, in an insulator $D_q{=}0$, while at a critical point $D_q$ is a
non-trivial function of $q$, implying wave function
multifractality. Splitting off the normal part, one defines the
anomalous dimensions $\Delta_q$,
\be
\label{e2.24}
\tau_q\equiv d(q-1)+\Delta_q,
\ee
which distinguish the critical point from the metallic phase and
determine the scale dependence of the wave function correlations.
Among them, $\Delta_2\equiv -\eta$ plays
the most prominent role, governing the spatial correlations of
the ``intensity'' $|\psi|^2$,
\be
L^{2d} \langle |\psi^2({\bf r})\psi^2({\bf r}')|\rangle
\sim (|{\bf r} - {\bf r}'|/L)^{-\eta}.
\label{e2.25}
\ee
Eq.~(\ref{e2.25}) can be obtained from (\ref{e2.23}) by using
that the wave function amplitudes become essentially uncorrelated at
$|\br-\br'|\sim L$. Scaling behavior of higher order
correlations,
$\langle|\psi^{2q_1}({\bf r_1})\psi^{2q_2}({\bf r_2})\ldots
\psi^{2q_n}({\bf r_n})|\rangle$, can be found in a similar way, e.g.
\begin{equation}
L^{d(q_1+q_2)}
\langle |\psi^{2q_1}({\bf r_1})\psi^{2q_2}({\bf r_2})|\rangle
\sim L^{-\Delta_{q_1}-\Delta_{q_2}}
(|{\bf r_1} - {\bf r_2}|/L)^{\Delta_{q_1+q_2}-\Delta_{q_1}-\Delta_{q_2}}.
\label{e2.25a}
\end{equation}
Correlations of different (close in energy) eigenfunctions
exhibit the same scaling  \cite{chalker90},
\be
\left.\begin{array}{l}
L^{2d} \langle |\psi_i^2({\bf r})\psi_j^2({\bf r}')|\rangle \\
L^{2d} \langle \psi_i({\bf r})\psi_j^*({\bf r})
\psi_i^*({\bf r}')\psi_j({\bf r}')\rangle
\end{array}  \right\}
\sim \left( \frac{|{\bf r} - {\bf r}'| }{L_\omega}\right)^{-\eta},
\label{e2.26}
\ee
where $\omega=\epsilon_i-\epsilon_j$,
$L_\omega\sim (\rho\omega)^{-1/d}$, $\rho$ is the density of states,
and $|{\bf r} - {\bf r}'| < L_\omega$.
For conventional classes, where the DOS is uncritical, the diffusion
propagator (\ref{e2.4}) scales in the same way.

In the field-theoretical language (Sec.~\ref{s2.2}), $\Delta_q$ are
the leading anomalous dimensions of the operators ${\rm Tr} (Q\Lambda)^q$
(or, more generally,  ${\rm Tr} (Q\Lambda)^{q_1} \ldots {\rm Tr}
(Q\Lambda)^{q_m} $ with $q_1{+}\ldots{+} q_m {=}q$) \cite{wegner80}. The
strong multifractal fluctuations of wave functions at criticality are
related to the fact that $\Delta_q{<}0$ for $q{>}1$, so that the
corresponding operators increase under RG. In this formalism, the
scaling of correlation functions [Eq.~(\ref{e2.25}) and its
generalizations] results from an operator product
expansion \cite{wegner85,duplantier91,mudry96}.

\subsubsection{Singularity spectrum $f(\alpha)$}
\label{s2.3.2}

The average IPR $\langle P_q\rangle$ are (up to the normalization
factor $L^d$) the moments of the distribution function ${\cal
  P}(|\psi|^2)$ of the eigenfunction intensities. The behavior
(\ref{e2.23}) of the moments corresponds to the intensity distribution
function of the form
\begin{equation}
\label{e2.27}
{\cal P}(|\psi^2|) \sim \frac{1}{|\psi^2|}L^{-d + f (-\frac{\ln |\psi^2|}{ \ln L} )}
\end{equation}
Indeed, calculating the moments $\langle |\psi^{2q}|\rangle$ with the
distribution  (\ref{e2.27}), one finds
\begin{equation}
\label{e2.28}
\langle P_q\rangle = L^d \langle |\psi^{2q}|\rangle \sim \int d\alpha\,
L^{-q\alpha+f(\alpha)}\ ,
\end{equation}
where we have introduced $\alpha=-\ln |\psi^2|/ \ln L$. Evaluation of
the integral by the saddle-point method (justified at
large $L$) reproduces
Eq.~(\ref{e2.23}), with the exponent $\tau_q$ related to the
singularity spectrum $f(\alpha)$ via the Legendre transformation,
\begin{equation}
\label{e2.29}
\tau_q=q\alpha-f(\alpha)\: , \qquad q=f'(\alpha)\: , \qquad \alpha =\tau'_q.
\end{equation}
The meaning of the function $f(\alpha)$ is as follows: it
is the fractal dimension of the set of those points ${\bf r}$ where the
eigenfunction intensity is $|\psi^2({\bf r})| \sim L^{-\alpha}$.
In other words, in a lattice version of the model the number of such points
scales as $L^{f(\alpha)}$ \cite{halsey86}.

General properties of $\tau_q$ and $f(\alpha)$ follow
from their definitions and the wave function normalization:

(i) $\tau_q$ is a non-decreasing, convex function ($\tau_q'\ge 0$,
$\tau_q''\le 0$ ), with $\tau_0=-d$, $\tau_1=0$;

(ii) $f(\alpha)$ is a convex function ($f''(\alpha)\le 0$) defined on
the semiaxis $\alpha\ge 0$ with a maximum at some point $\alpha_0$
(corresponding to $q=0$ under the Legendre transformation) and
$f(\alpha_0)=d$. Further, for the point $\alpha_1$ (corresponding to
$q=1$) we have $f(\alpha_1) = \alpha_1$ and $f'(\alpha_1)=1$.

If one formally defines $f(\alpha)$ for a metal, it will be
concentrated in a single point $\alpha=d$, with $f(d)=d$ and
$f(\alpha)= - \infty$ otherwise. On the other hand, at criticality
this ``needle'' broadens and the maximum shifts to a position
$\alpha_0>d$, see Fig.~\ref{fig1}.

\subsubsection{Symmetry of the multifractal spectra}
\label{s2.3.4}

\begin{figure}[ht]
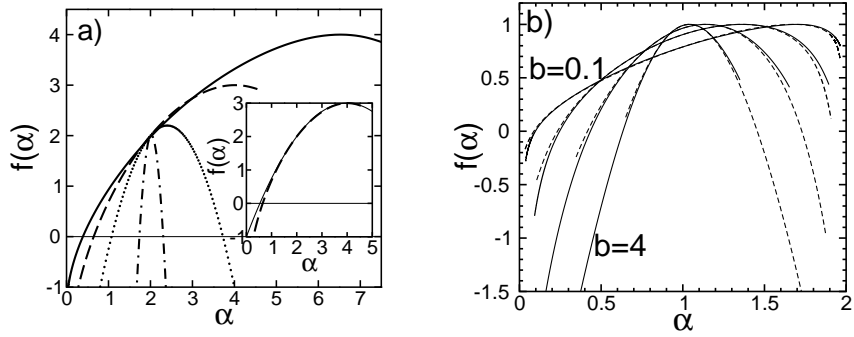
%
\begin{center}
\parbox{5.5cm}{
\includegraphics[width=5.3cm,clip]{fa.2+epsD3D4.pp.eps} }
\hspace*{0.5cm}
\parbox{5.5cm}{
\includegraphics[width=5.3cm,clip]{fa.prbm.pp.eps} }
\caption{Multifractality at Anderson transitions.
{\it a)} Singularity spectrum $f(\alpha)$ in $d=2+\epsilon$ for
$\epsilon=0.01$ and $\epsilon=0.2$ (analytical), as well as 3D and
4D (numerical). With increasing $d$ the spectrum gets broader,
implying stronger multifractality. Inset: comparison between
$f(\alpha)$ for 3D and the one-loop result of the $2+\epsilon$
expansion with $\epsilon=1$ (solid).\cite{mildenberger02}
{\it b)} Singularity spectrum $f(\alpha)$ for the PRBM
model. Evolution from weak to strong multifractality with decreasing
parameter $b$ is evident.
Dashed lines represent $f(2-\alpha)+\alpha-1$,
  demonstrating the validity of Eq.~(\ref{e2.33}).
\protect\cite{mildenberger07a}.}
\label{fig1}
  \end{center}
\vspace*{-0.5cm}
\end{figure}

As was recently shown\cite{mirlin06}, the multifractal exponents for the
Wigner-Dyson classes satisfy an exact symmetry relation
\be
\label{e2.32}
\Delta_q = \Delta_{1-q}\ ,
\ee
connecting  exponents with $q<1/2$ (in particular, with negative
$q$) to those with $q>1/2$. In terms of the singularity spectrum, this
implies
\be
\label{e2.33}
f(2d-\alpha) = f(\alpha) + d -\alpha.
\ee
The analytical derivation of Eqs.~(\ref{e2.32}), (\ref{e2.33})
is based on the supersymmetric $\sigma$-model;
it has been confirmed by numerical simulations on the
PRBM model \cite{mirlin06,mildenberger07a} (see Fig.~\ref{fig1}b) and 2D
Anderson transition of the symplectic class
\cite{mildenberger07,obuse07}.

\subsubsection{Dimensionality dependence of multifractality}
\label{s2.3.5a}


Let us analyze the evolution\cite{mildenberger02}
from the weak-multifractality regime in
$d=2+\epsilon$
dimensions to the strong multifractality at $d\gg 1$.

In $2+\epsilon$ dimensions with $\epsilon\ll 1$ the multifractality exponents
can be obtained within the $\epsilon$-expansion, Sec.~\ref{s2.2.2}. The 4-loop
results for the orthogonal and unitary symmetry classes read \cite{wegner87}
\bea
 \Delta_q^{(O)} &=& q(1-q)\epsilon + \frac{\zeta(3)}{  4}q(q-1)(q^2-q+1)\epsilon^4 + O(\epsilon^5);
\label{e2.42}\\
\Delta_q^{(U)} &=& q(1-q)(\epsilon/2)^{1/2} -
\frac{3}{8}q^2(q-1)^2\zeta(3)\epsilon^2
+ O(\epsilon^{5/2}).
\label{e2.43}
\eea
Keeping only the leading term on the r.h.s. of Eqs.~(\ref{e2.42})
and (\ref{e2.43}), we get the one-loop approximation for $\tau_q$
which is of parabolic form.

Numerical simulations \cite{mildenberger02} of the wave function
statistics in 3D and 4D (Fig.~\ref{fig1}a)
have shown a full qualitative agreement with
analytical predictions, both in the form of multifractal spectra and
in the shape of the IPR distribution. Moreover,
the one-loop result of the $2{+}\epsilon$ expansion with $\epsilon{=}1$
describes the 3D singularity spectrum with a remarkable accuracy (though with
detectable deviations). In particular, the position of the maximum,
$\alpha_0{=}4.03{\pm}0.05$, is very close to its value $\alpha_0{=}d{+}\epsilon$
implied by one-loop approximation. As expected, in 4D the deviations from
parabolic shape are much more pronounced and $\alpha_0{=}6.5{\pm}0.2$ differs
noticeably from 6.

The simulations\cite{mildenberger02}
also show that fractal dimensions $D_q\equiv
\tau_q/(q-1)$ with
$q\gtrsim 1$ decrease with increasing $d$. As an example, for $q=2$ we have
$D_2\simeq 2-2\epsilon$ in $2+\epsilon$ dimensions, $D_2=1.3\pm 0.05$ in 3D,
and $D_2=0.9\pm 0.15$ in 4D. This confirms the expectation based on the
Bethe-lattice results  (Sec.~\ref{s2.4}) that $\tau_q\to 0$ at $d\to\infty$ for
$q>1/2$. Such a behavior of the multifractal exponents is a manifestation of a
very sparse character of critical eigenstates at $d\gg 1$, formed by rare
resonance spikes. In combination with the  relation (\ref{e2.32})
this implies the limiting form of the multifractal spectrum at $d\to\infty$,
\be
\label{e2.46}
\tau_q=\left\{ \begin{array}{ll}
0\ , & \qquad q\ge 1/2 \\
2d(q-1/2)\ , & \qquad q\le 1/2\ .
\end{array}
\right.
\ee
This corresponds to $f(\alpha)$ of the form
\be
\label{e2.47}
f(\alpha) = \alpha/2\ , \qquad 0<\alpha<2d\ ,
\ee
dropping to $-\infty$ at the boundaries of the interval $[0,2d]$.
It was argued\cite{mildenberger02} that the way the
multifractality spectrum approaches this limiting form with increasing $d$ is
analogous to the behavior found\cite{evers00} in the PRBM model with $b\ll 1$.

\subsubsection{Surface vs. bulk multifractality}
\label{s2.3.7}

Recently, the concept of wave function multifractality was
extended \cite{subramaniam06} to
the surface of a system at an Anderson
transition. It was shown that fluctuations of
critical wave functions at the surface are characterized by a new set of
exponent $\tau_q^{\rm s}$ (or, equivalently, anomalous exponents
$\Delta_q^{\rm s})$ independent from their bulk counterparts,
\begin{eqnarray}\label{tau-q-s}
&& L^{d-1} \langle |\psi({\bf r})|^{2q} \rangle \sim L^{-\tau_q^{\rm
s}}, \\
&& \tau_q^{\rm s} = d(q-1) + q\mu + 1 + \Delta_q^{\rm s}.
\label{delta-q-s}
\end{eqnarray}
Here $\mu$ is introduced for generality, in order to account for
a possibility of  non-trivial scaling of the average
value, $\langle|\psi({\bf r})|^2\rangle
\propto L^{-d-\mu}$,  at the boundary in unconventional symmetry
classes. For the Wigner-Dyson classes, $\mu=0$.
The normalization factor
$L^{d-1}$ is chosen such that Eq.~(\ref{tau-q-s}) yields the
contribution of the surface to the IPR
$\langle P_q\rangle = \langle \int d^d {\bf r} |\psi({\bf
r})|^{2q} \rangle $. The exponents $\Delta_q^{\rm s}$ as defined in
Eq.~(\ref{delta-q-s})  vanish in a metal and
govern statistical fluctuations of wave
functions at the boundary, $\langle |\psi({\bf r})|^{2q} \rangle /
\langle |\psi({\bf r})|^2 \rangle^q \sim L^{-\Delta_q^{\rm s}}$,
as well as their spatial correlations, e.g. $L^{2(d+\mu)} \langle
|\psi^2({\bf r})\psi^2({\bf r}')|\rangle \sim (|{\bf r} - {\bf
r}'|/L)^{\Delta_2^{\rm s}}$.

Wave function fluctuations are much stronger at the edge than in the bulk.
As a result, surface exponents are important even if
one performs a multifractal analysis for the whole sample, without
separating it into ``bulk'' and ``surface'', despite the fact that the
weight of surface points is down by a factor $1/L$.

The boundary multifractality was
explicitly studied, analytically as well as numerically, for a variety
of critical systems, including weak multifractality in 2D and
$2+\epsilon$ dimensions,
the 2D spin quantum Hall transition \cite{subramaniam06},
the Anderson transition in a 2D system with spin-orbit coupling
\cite{obuse07}, and the PRBM model \cite{mildenberger07a}.
The notion of surface multifractality was further
generalized\cite{obuse07} to a corner of a critical system.

\subsection{Additional comments}
\label{s2.3.additional}

\begin{enumerate}

\item[(i)]
For the lack of space we do not discuss the issues of IPR
distributions at criticality and the role of ensemble averaging,
as well as possible singularities in multifractal spectra, see
the review \refcite{evers08}.

\item[(ii)]
Recently, an impressive progress was achieved in experimental studies
of Anderson transitions in various
systems.\cite{morgenstern08,chabe08,hu08,faez09,richardella10}
The developed experimental
techniques permit spatially resolved investigation of wave functions,
thus paving the way to experimental study of multifractality. While the
obtained multifractal spectra differ numerically from theoretical expectations
(possibly because the systems were not exactly at criticality, or, in
the case of electronic systems, pointing to importance of
electron-electron interaction), the experimental advances seem very
promising.

\item[(iii)]
Recent theoretical work\cite{feigelman10}
explains the properties of a superconductor-insulator
transition observed in a class of disordered films in terms of
multifractality of electronic wave functions.

\end{enumerate}

\subsection{Anderson transition in $d=\infty$: Bethe lattice}
\label{s2.4}

The Bethe lattice (BL) is a tree-like lattice with a fixed coordination
number. Since the number of sites at a distance $r$ increases
exponentially with $r$ on the BL, it effectively corresponds to
the limit of high dimensionality $d$.
The BL models are the closest existing analogs of
the mean-field theory for the case of the Anderson transition.
The Anderson tight-binding model (lattice version of
Eqs.~(\ref{e2.2a}), (\ref{e2.2b})) on the BL was studied for the first
time in Ref.~\refcite{abou-chacra73}, where the existence of the
metal-insulator transition was proven and the position of the mobility
edge was determined.
Later, the BL versions of the
$\sigma$-model (\ref{e2.12})
\cite{efetov85,zirnbauer86}
and of the tight-binding model \cite{mirlin91} were studied within the
supersymmetry formalism, which allowed to determine the critical
behavior. It was found that the localization length diverges in the
way usual for BL models, $\xi\propto |E-E_c|^{-1}$, where $E$ is a
microscopic parameter driving the transition. When reinterpreted
within the effective-medium approximation  \cite{efetov90,fyodorov92},
this yields the conventional mean-field value of the
localization length exponent, $\nu=1/2$. On the other hand, the
critical behavior of other observables is very peculiar. The
inverse participation ratios $P_q$ with $q>1/2$ have a finite limit at
$E\to E_c$ when the critical point is approached from the localized
phase and then jump to zero. By comparison with the scaling formula,
$P_q\propto \xi^{-\tau_q}$, this can be interpreted as $\tau_q=0$ for
all $q\ge 1/2$. Further, in the delocalized phase the diffusion
coefficient vanishes exponentially when the critical point is
approached,
\begin{eqnarray}
\label{e2.54}
&& D \propto \Omega^{-1} \ln^3 \Omega\ ; \qquad
 \Omega \sim \exp\{{\rm const}\:|E-E_c|^{-1/2}\},
\end{eqnarray}
which can be thought as corresponding to the infinite value,
$s=\infty$, of the critical index $s$. The distribution of
the LDOS $v\equiv\rho(r)/\langle\rho\rangle$ (normalized to its
average value for convenience) was found to be of the form
\be
\label{e2.56}
{\cal P}(v) \propto \Omega^{-1/2} v^{-3/2}\ , \qquad \Omega^{-1} \ll v
\ll \Omega\ ,
\ee
and exponentially small outside this range.
Equation (\ref{e2.56}) implies for the moments of the LDOS:
\be
\label{e2.57}
\langle v^q\rangle \propto \Omega^{|q-1/2|-1/2}.
\ee
The physical reason for the unconventional critical behavior was
unraveled in Ref.~\refcite{mirlin94}.
It was shown that the exponential largeness of
$\Omega$ reflects the spatial structure of the BL: the ``correlation
volume'' $V_\xi$ (number of sites within a distance $\xi$ from the
given one) on such a lattice is exponentially large. On the other
hand, for any finite dimensionality $d$ the correlation volume has a
power-law behavior, $V_d(\xi)\propto \xi^d \propto |E-E_c|^{\nu d}$,
where $\nu\simeq 1/2$ at large $d$. Thus, the scale $\Omega$ cannot
appear for finite $d$ and, assuming some matching between the BL and
large-$d$ results, will be replaced by $V_d(\xi)$. Then
Eq.~(\ref{e2.57}) yields the following high-$d$ behavior of the
anomalous exponents $\Delta_q$ governing the scaling of the LDOS
moments (Sec.~\ref{s2.3}),
\be
\label{e2.58}
\Delta_q \simeq d(1/2 - |q-1/2|)\ ,
\ee
or, equivalently, the results (\ref{e2.46}), (\ref{e2.47}) for the
multifractal spectra $\tau_q$, $f(\alpha)$. These formulas describe
the strongest possible multifractality.

The critical behavior of the conductivity, Eq.~(\ref{e2.54}), is
governed by the same exponentially large factor $\Omega$. When it is replaced
by the correlation volume $V_d(\xi)$, the power-law
behavior at finite $d\gg 1$ is recovered,
$\sigma\propto|E-E_c|^s$ with $s\simeq d/2$.
The result for the exponent $s$ agrees (within its accuracy,
i.e. to the leading order in $d$) with the scaling relation
$s=\nu(d-2)$.

\section{Symmetries of disordered systems}
\label{s:SymDisSys}

In this section we briefly review the symmetry classification of
disordered systems based on the relation to the classical symmetric
spaces, which was established in Refs. \refcite{altland97,zirnbauer96b}.

\subsection{ Wigner-Dyson classes}
\label{s4.1}

The random matrix theory (RMT) was introduced into physics
by Wigner\cite{wigner51}. Developing
Wigner's ideas, Dyson\cite{dyson62} put forward a
classification scheme of ensembles of random Hamiltonians.
This scheme takes into account the invariance of the system under time
reversal and spin rotations, yielding three symmetry classes: unitary,
orthogonal and symplectic.
If the time-reversal invariance ($T$) is broken, the Hamiltonians are just
arbitrary Hermitian matrices,
\be
\label{e4.1}
H=H^\dagger\ ,
\ee
with no further
constraints. This set of matrices is invariant with respect to
rotations by unitary matrices; hence the name ``unitary ensemble''.
In this situation, the presence or absence of spin rotation invariance ($S$)
is not essential: if the spin is conserved, $H$ is simply a
spinless unitary-symmetry Hamiltonian times the unit matrix in the spin space.
In the RMT one considers most frequently an ensemble of matrices with
independent, Gaussian-distributed random entries -- the
Gaussian unitary ensemble (GUE). While disordered systems have
much richer physics than the Gaussian ensembles, their
symmetry classification is inherited from the RMT.

Let us now turn to the systems with preserved time-reversal
invariance. The latter is represented by an antiunitary operator,
$T=KC$, where $C$ is the operator of complex conjugation and $K$ is
unitary. The time-reversal invariance thus implies
$H=KH^{\rm T}K^{-1}$ (we used the Hermiticity, $H^* = H^{\rm T}$).
Since acting twice with $T$ should leave the physics
unchanged, one infers that $K^*K = p$, where $p=\pm 1$.
As was shown by Wigner, the two cases correspond to systems with
integer ($p=+1$) and half-integer ($p=-1$) angular momentum.
If $p=1$, a representation can be chosen where $K=1$, so that
\be
\label{e4.2}
H=H^{\rm T}\ .
\ee
The set of Hamiltonians thus spans the space of real
symmetric matrices in this case. This is the orthogonal symmetry
class; its representative is the Gaussian orthogonal ensemble (GOE).
For disordered electronic systems this class is realized when spin
is conserved, as the Hamiltonian then reduces to that for spinless
particles (times unit matrix in the spin space).

If  $T$ is preserved but $S$ is broken, we have $p=-1$. In the
standard representation, $K$ is then realized by the second Pauli
matrix, $K =i\sigma_y$, so that the Hamiltonian satisfies
\be
\label{e4.3}
H = \sigma_y H^{\rm T} \sigma_y.
\ee
It is convenient to split the $2N \times 2N$
Hamiltonian in $2\times 2$ blocks (quaternions) in spin
space. Each of them then is of the form $q = q_0\sigma_0 +
iq_1\sigma_x +i q_2\sigma_y + i q_3\sigma_z$ (where $\sigma_0$ is the
unit matrix and $\sigma_{x,y,z}$ the Pauli matrices), with
real $q_\mu$, which defines a real quaternion. This set of
Hamiltonians is invariant with respect to the group of unitary
transformations
conserving $\sigma_y$, $U \sigma_y U^T = \sigma_y$, which is the
symplectic group $\text{Sp}(2N)$. The corresponding symmetry class is
thus called symplectic, and its RMT representative is the Gaussian
symplectic ensemble (GSE).

\subsection{ Relation to symmetric spaces}
\label{s4.2}

Before discussing the relation to the families of symmetric
spaces, we briefly remind the reader how the latter are constructed
\cite{helgason78,caselle04}.
Let G be one of the compact Lie groups $\text{SU}(N)$,
$\text{SO}(N)$, $\text{Sp}(2N)$, and $\mathfrak{g}$ the corresponding
Lie algebra. Further, let $\theta$ be an involutive automorphism
$\mathfrak{g} \to \mathfrak{g}$ such that $\theta^2=1$ but $\theta$ is
not identically equal to unity. It is clear that $\theta$ splits
$\mathfrak{g}$ in two complementary subspaces,
$\mathfrak{g} = \mathfrak{K} \oplus \mathfrak{P}$, such that
$\theta(X)=X$ for $X\in\mathfrak{K}$ and
$\theta(X)=-X$ for $X\in\mathfrak{P}$. It is easy to see that
the following Lie algebra multiplication relations holds:
\be
\label{e4.4}
[\mathfrak{K},\mathfrak{K}]\subset \mathfrak{K},\qquad
[\mathfrak{K},\mathfrak{P}]\subset \mathfrak{P},\qquad
[\mathfrak{P},\mathfrak{P}]\subset \mathfrak{K}.
\ee
This implies, in particular,  that $\mathfrak{K}$ is a subalgebra,
whereas $\mathfrak{P}$ is not. The coset space $G/K$ (where $K$ is the
Lie group corresponding to $\mathfrak{K}$) is then a compact symmetric
space. The tangent space to $G/K$ is  $\mathfrak{P}$.
One can also construct an associated non-compact space. For this
purpose, one first defines the Lie algebra
$\mathfrak{g}^*=\mathfrak{K} \oplus i\mathfrak{P}$, which differs from
$\mathfrak{g}$ in that the elements in $\mathfrak{P}$ are multiplied
by $i$. Going to the corresponding group and dividing $K$ out, one
gets a non-compact symmetric space $G^*/K$.

The groups $G$ themselves are also symmetric spaces and can be viewed
as coset spaces $G\times G/G$. The corresponding non-compact space is
$G^\mathbb{C}/G$, where $G^\mathbb{C}$ is the complexification of $G$
(which is obtained by taking the Lie algebra $\mathfrak{g}$, promoting
it to the algebra over the field of complex numbers, and then
exponentiating).

The connection with symmetric spaces is now established in the following
way \cite{altland97,zirnbauer96b}. Consider first the unitary symmetry
class. Multiplying a Hamiltonian matrix by $i$, we get an
antihermitean matrix $X=iH$. Such matrices form the Lie algebra
$\mathfrak{u}(N)$. Exponentiating it, one gets the Lie group
$\text{U}(N)$, which is the compact symmetric space of class A in
Cartan's classification.
For the orthogonal class, $X=iH$ is purely
imaginary and symmetric. The set of such matrices is a linear complement
$\mathfrak{P}$ of the algebra $\mathfrak{K}=\mathfrak{o}(N)$ of imaginary
antisymmetric matrices  in the algebra $\mathfrak{g}=\mathfrak{u}(N)$
of antihermitean matrices. The corresponding symmetric space is
$G/K = \text{U}(N)/\text{O}(N)$, which  is termed AI in Cartan's
classification.   For the symplectic ensemble the same consideration
leads to the symmetric space $\text{U}(N)/\text{Sp}(N)$, which is the
compact space of the class AII. If we don't multiply $H$ by $i$ but
instead proceed with $H$ in the analogous way, we end up with
associated non-compact spaces $G^*/K$. To summarize, the linear space
$\mathfrak{P}$ of Hamiltonians can be considered as a tangent space to
the compact $G/K$ and non-compact $G^*/K$ symmetric spaces of the
appropriate symmetry class.

\begin{sidewaystable*}
\tbl{Symmetry classification of disordered systems. First column:
  symbol for the symmetry class of the Hamiltonian. Second column:
  names of the corresponding RMT. Third column: presence (+) or
  absence ($-$) of the time-reversal (T) and spin-rotation (S)
  invariance. Fourth and fifth columns: families of the compact and
  non-compact symmetric spaces of the corresponding symmetry class.
The Hamiltonians span the tangent space to these symmetric spaces.
Sixth column: symmetry class of the $\sigma$-model; the first symbol
corresponds to the non-compact (``bosonic'') and the second to the
compact (``fermionic'') sector of the base of the $\sigma$-model
manifold. The compact component ${\cal M}_F$ (which is particularly
important for theories with non-trivial topological properties) is
explicitly given in the last column. From Ref.~\refcite{evers08}}
{\begin{tabular}{|c|c|c|c|c|c|c|}
\hline
Ham. & RMT & T \ \  S & compact & non-compact & $\sigma$-model &
$\sigma$-model compact
\\
 class &    &  & symmetric space & symmetric space & B$|$F & sector ${\cal M}_F$
\\
\hline
\multicolumn{7}{l}{Wigner-Dyson classes}\\
\hline
A & GUE  & $-$ \ \  $\pm$ & $\text{U}(N){\times}\text{U}(N)/\text{U}(N)
\equiv \text{U}(N) $ & $\text{GL}(N,\mathbb{C})/\text{U}(N)$ & AIII$|$AIII
& $\text{U}(2n)/\text{U}(n){\times}\text{U}(n)$
\\
\hline
AI & GOE & $+$ \ \ $+$ &  $\text{U}(N)/\text{O}(N)$ &
 $\text{GL}(N,\mathbb{R})/\text{O}(N)$
 & BDI$|$CII & $\text{Sp}(4n)/\text{Sp}(2n){\times}\text{Sp}(2n)$
\\
\hline
AII & GSE & $+$ \ \ $-$ & $\text{U}(2N)/\text{Sp}(2N)$ &
 $\text{U}^*(2N)/\text{Sp}(2N)$ &  CII$|$BDI &
$\text{O}(2n)/\text{O}(n){\times}\text{O}(n)$
\\
\hline
\multicolumn{7}{l}{chiral classes}\\
\hline
AIII & chGUE  &  $-$ \ \  $\pm$ &
$\text{U}(p+q)/\text{U}(p){\times}\text{U}(q)$ &
$\text{U}(p,q)/\text{U}(p){\times}\text{U}(q)$
& A$|$A & $\text{U}(n)$
\\
\hline
BDI & chGOE & $+$ \ \ $+$ &
$\text{SO}(p+q)/\text{SO}(p){\times}\text{SO}(q)$ &
$\text{SO}(p,q)/\text{SO}(p){\times}\text{SO}(q)$
& AI$|$AII &  $\text{U}(2n)/\text{Sp}(2n)$
\\
\hline
CII & chGSE & $+$ \ \ $-$ &
$\text{Sp}(2p+2q)/\text{Sp}(2p){\times}\text{Sp}(2q)$ &
$\text{Sp}(2p,2q)/\text{Sp}(2p){\times}\text{Sp}(2q)$
& AII$|$AI &  $\text{U}(n)/\text{O}(n)$
\\
\hline
\multicolumn{7}{l}{Bogoliubov - de Gennes classes}\\
\hline
C &  & $-$ \ \  $+$ &
$\text{Sp}(2N){\times}\text{Sp}(2N)/\text{Sp}(2N) \equiv \text{Sp}(2N)$
&  $\text{Sp}(2N,\mathbb{C})/\text{Sp}(2N)$  &
DIII$|$CI & $\text{Sp}(2n)/\text{U}(n)$
\\
\hline
CI &  &  $+$ \ \  $+$ & $\text{Sp}(2N)/\text{U}(N)$ &
$\text{Sp}(2N,\mathbb{R})/\text{U}(N)$  &
D$|$C & $\text{Sp}(2n)$
\\
\hline
BD & &  $-$ \ \  $-$ &
$\text{SO}(N){\times}\text{SO}(N)/\text{SO}(N)\equiv \text{SO}(N)$ &
$\text{SO}(N,\mathbb{C})/\text{SO}(N)$  &
CI$|$DIII  &  $\text{O}(2n)/\text{U}(n)$
\\
\hline
DIII & & $+$ \ \  $-$ &  $\text{SO}(2N)/\text{U}(N)$ &
 $\text{SO}^*(2N)/\text{U}(N)$ & C$|$D  & $\text{O}(n)$
\\
\hline
\end{tabular} }
\label{t4.1}
\end{sidewaystable*}

This correspondence is summarized in Table~\ref{t4.1}, where the first
three rows correspond to the Wigner-Dyson classes, the next three to
the chiral classes (Sec.~\ref{s4.3}) and last four to the
Bogoliubov-de Gennes classes  (Sec.~\ref{s4.4}). The last two columns
of the table specify the symmetry of the corresponding $\sigma$-model.
In the supersymmetric formulation, the base of the $\sigma$-model
target space $\cal{M}_B\times\cal{M}_F$ is the product of a
non-compact symmetric space $\cal{M}_B$ corresponding to the bosonic
sector and a compact (``fermionic'') symmetric space $\cal{M}_F$.
(In the replica formulation, the space is $\cal{M}_B$ for bosonic
or $\cal{M}_F$ for fermionic replicas, supplemented with the
limit $n\to 0$.)
The Cartan symbols for these symmetric spaces are given in the sixth
column, and the compact components  $\cal{M}_F$ are listed in the last
column. It should be stressed that the symmetry classes of
$\cal{M}_B$ and $\cal{M}_F$ are different from the symmetry class of
the ensemble (i.e. of the Hamiltonian) and in most cases are also
different from each other.
Following the common convention, when we refer to a system as
belonging to a particular class,
we mean the symmetry class of the Hamiltonian.

It is also worth emphasizing that the orthogonal groups appearing
in the expressions for  $\cal{M}_F$
are $\text{O}(N)$ rather than $\text{SO}(N)$. This difference
(which was irrelevant when we were discussing the symmetry of the
Hamiltonians, as it does not affect the tangent space) is
important here, since it influences topological properties of the
manifold. As we will detail in Sec.~\ref{s6},\ref{s8}, the
topology of the $\sigma$-model target space often affects the
localization properties of the theory in a crucial way.

\subsection{ Chiral classes}
\label{s4.3}

The Wigner-Dyson classes are the only allowed if one looks for a
symmetry that is translationally invariant in energy, i.e. is not
spoiled by adding a constant to the Hamiltonian. However, additional
discrete symmetries may arise at some particular value of energy
(which can be chosen to be zero without loss of generality), leading
to novel symmetry classes. As the vicinity of a special point in the
energy space governs the physics in many cases (i.e. the band center
in  lattice models at half filling, or zero energy in gapless
superconductors), these ensembles  are of large interest. They can be
subdivided into two groups -- chiral and Bogoliubov - de Gennes
ensembles -- considered here and in Sec.~\ref{s4.4}, respectively.

The chiral ensembles appeared in both contexts of particle physics
and physics of disordered electronic systems about fifteen years
ago \cite{gade93,gade91,slevin93,verbaarschot93}. The corresponding
Hamiltonians have the form
\begin{equation}
\label{e4.5}
   { H } = \left( \begin{array}{cc}
             0           &    h \\
             h^{\dagger} &    0
             \end{array}
             \right)\ ,
\end{equation}
i.e. they possess the symmetry
\be
\label{e4.6}
\tau_z H \tau_z = - H \ ,
\ee
where $\tau_z$ is the third Pauli matrix in a certain ``isospin''
space. In the condensed matter context, such ensembles arise, in
particular, when one considers a tight-binding model on a bipartite
lattice with randomness in hopping matrix elements only. In this case,
$H$ has the block structure (\ref{e4.5}) in the sublattice space.

In addition to the chiral symmetry, a system may possess
time reversal and/or spin-rotation invariance.
In full analogy with the
Wigner-Dyson classes, \ref{s4.1}, one gets therefore three chiral
classes (unitary, orthogonal, and symplectic). The corresponding
symmetric spaces, the Cartan notations for symmetry classes, and the
$\sigma$-model manifolds are given in the rows 4--6 of the
Table~\ref{t4.1}.

\subsection{ Bogoliubov - de Gennes classes}
\label{s4.4}

The Wigner-Dyson and chiral classes
do not exhaust all possible symmetries of disordered electronic
systems\cite{altland97}. The remaining four classes arise most naturally in
superconducting systems. The quasiparticle dynamics in such systems can be
described by the Bogoliubov-de Gennes Hamiltonian of the form
\begin{equation}
\label{e4.7}
\hat H = \sum_{\alpha\beta}^{N} h_{\alpha\beta} c^{\dagger}_{\alpha} c_{\beta}
+ \frac{1}{2} \sum_{\alpha\beta}^{N} \left( \Delta_{\alpha\beta}
c^\dagger_\alpha c^{\dagger}_\beta -
  \Delta_{\alpha\beta}^* c_\alpha c_\beta \right),
\end{equation}
where $c^\dagger$ and $c$ are fermionic creation and annihilation operators,
and the $N\times N$ matrices $h$, $\Delta$ satisfy $h=h^{\dagger}$ and
$\Delta^T = -\Delta$, in view of hermiticity.
Combining $c^{\dagger}_{\alpha},c_{\alpha}$ in a spinor
$\psi_{\alpha}^{\dagger}=(c_{\alpha}^{\dagger},c_{\alpha})$, one gets
a matrix representation of the Hamiltonian,
$\hat H = \psi^\dagger H \psi$, where
\begin{equation}
\label{e4.8}
      H = \left( \begin{array}{cc}
             h           &    \Delta\\
             -\Delta^*   &    -h^T
             \end{array}
             \right), \qquad h = h^\dagger\ ,\ \ \Delta=-\Delta^T\ .
\end{equation}
The minus signs in the definition of $H$ result form the fermionic
commutation relations between $c^{\dagger}$ and $c$.
The Hamiltonian structure (\ref{e4.8}) corresponds to
the condition
\be
\label{e4.9}
H=-\tau_x H^T \tau_x
\ee
(in addition to the Hermiticity $H=H^\dagger$), where $\tau_x$ is the
Pauli matrix in the particle-hole space.  Alternatively, one can
perform a unitary rotation of the basis,
defining $\tilde{H}= g^\dagger H g$ with $g =
(1+i\tau_x)/\sqrt{2}$.
In this basis, the defining condition of class D becomes $\tilde{H} =
- \tilde{H}^T$,  so that $\tilde{H}$ is pure imaginary.
The matrices $X=iH$ thus form the Lie algebra $\mathfrak{so}(2N)$,
corresponding to the Cartan class D. This symmetry class described
disordered superconducting systems in the absence of other symmetries.

Again, the symmetry class will be changed if the time reversal and/or
spin rotation invariance are present. The difference with respect to
the Wigner-Dyson and chiral classes is that now one gets four
different classes rather than three. This is because the spin-rotation
invariance has an impact even in the absence of time-reversal
invariance, since it combines  with the particle-hole symmetry in a
non-trivial
way. Indeed, if the spin is conserved, the Hamiltonian has the form
  \begin{equation}
\label{e4.10}
    \hat H = \sum_{ij}^{N} \left[h_{ij}(c_{i\uparrow}^\dagger c_{j\uparrow} -
    c_{j\downarrow} c^\dagger_{i,\downarrow})
    + \Delta_{ij}
    c^{\dagger}_{i,\uparrow}c^{\dagger}_{j,\downarrow}
    + \Delta^*_{ij} c_{i\downarrow} c_{j\uparrow}\right],
    \end{equation}
where $h$ and $\Delta$ are $N{\times}N$ matrices satisfying
$h=h^\dagger$ and $\Delta=\Delta^T$.
Similar to (\ref{e4.8}), we can introduce the spinors
$\psi_{i}^{\dagger} = (c_{i\uparrow}^\dagger, c_{i\downarrow})$ and
obtain the following matrix form of the Hamiltonian
\begin{equation}
\label{e4.11}
   { H } = \left(
     \begin{array}{cc}
             { h } & { \Delta} \\
             { \Delta}^* & { -h}^{T}
             \end{array}
             \right), \qquad h = h^\dagger\ ,\ \ \Delta=\Delta^T\ .
\end{equation}
It exhibits a symmetry property
\begin{equation}
\label{e4.12}
  H =   - \tau_{y} H^T \tau_y.
\end{equation}
The matrices $H=iX$ now form the Lie algebra $\mathfrak{sp}(2N)$,
which is the symmetry class C.

If the time reversal invariance is present, one gets two more classes
(CI and DIII). The symmetric spaces for the Hamiltonians and the
$\sigma$-models corresponding to the Bogoliubov--de Gennes classes are
given in the last four rows of the Table~\ref{t4.1}.

The following comment is in order here.
Strictly speaking, one should distinguish between the orthogonal group
$\mathrm{SO}(N)$ with even and odd $N$, which form different Cartan
classes: $\mathrm{ SO}(2N)$ belongs to class D, while  $\mathrm{ SO}(2N{+}1)$
to class B. In the conventional situation of a disordered
superconductor, the matrix size is even  due to the
particle-hole space doubling, see Sec.~\ref{s4.4}. It was found, however,
that the class B can arise in $p$-wave vortices \cite{ivanov02}.
In the same sense, the class DIII should be
split in DIII-even and DIII-odd; the last one represented by the
symmetric space $\text{SO}(4N+2)/\text{U}(2N+1)$ can appear in
vortices in the presence of time-reversal symmetry.

\subsection{Perturbative RG for $\sigma$-models of different symmetry classes}
\label{s4.6}

Perturbative $\beta$-functions for $\sigma$-models on all the types of
symmetric spaces were in fact calculated \cite{hikami81,wegner88}
long before the
physical significance of the chiral and Bogoliubov-de Gennes classes
has been fully appreciated. These results are important
for understanding the behavior of systems of different
symmetry classes in 2D. (We should emphasize once more, however, that this
does not give a complete information about all possible types of
criticality since the latter can be crucially affected by additional
terms of topological character in the $\sigma$-model, see
Sec.~\ref{s6}, \ref{s8} below.)

One finds that in
the classes A, AI, C, CI the $\beta$-function is negative in 2D in the
replica limit (at least, for small $t$). This indicates that normally
all states are localized in such systems in 2D. (This conclusion can
in fact be changed in the presence of topological or Wess-Zumino
terms, Sec.~\ref{s6.1}.) Above 2D, these systems undergo the Anderson
transition that can be studied within the $2+\epsilon$ expansion,
Sec.~\ref{s2.2.2}.  For the classes AIII, BDI, and CII (chiral
unitary, orthogonal and symplectic classes, respectively) the
$\beta(t)\equiv 0$ in 2D, implying a line of fixed
points.
Finally, in the classes AII, D, and DIII the
$\beta$-function is positive at small $t$, implying the existence of a
metal-insulator transition at strong coupling in 2D.

\section{Criticality in 2D}
\label{s6}

\subsection{Mechanisms of criticality in 2D}
\label{s6.1}

As was discussed in Sec.~\ref{s2.2.2}, conventional Anderson
transitions in the orthogonal and unitary symmetry classes take place
only if the dimensionality is $d>2$, whereas in 2D all states are
localized. It is, however, well understood by now that there is a rich
variety of mechanisms that lead to emergence of criticality in 2D
disordered systems.\cite{fendley00}
Such 2D critical points have been found to exist
for 9 out of 10 symmetry classes, namely, in all classes
except for the orthogonal class AI.
A remarkable peculiarity of 2D critical points is that the critical
conductance $g_*$ is at the same time the {\it critical conductivity}.
We now list and briefly describe the mechanisms for
the emergence of criticality.

\subsubsection{Broken spin-rotation invariance: Metallic phase}
\label{s6.1.1}

We begin with the mechanism that has been already mentioned in
Sec.~\ref{s2.2.2} in the context of the Wigner-Dyson symplectic class
(AII). In this case the $\beta$-function  [(\ref{e2.20}) with
$\epsilon=0$] is positive for not too large $t$ (i.e. sufficiently
large conductance), so that the system is metallic ($t$ scales to zero
under RG). On the other hand, for strong disorder (low $t$) the system
is an insulator, as usual, i.e. $\beta(t)<0$. Thus, $\beta$-function
crosses zero at some $t_*$, which is a point of the Anderson
transition.

This mechanism  (positive $\beta$-function and, thus, metallic
phase at small $t$, with a transition at some $t_*$) is also realized in
two of Bogoliubov-de Gennes classes -- D and DIII.
All these
classes correspond to systems with broken spin-rotation invariance.
The unconventional sign of the $\beta$-function in these classes,
indicating weak antilocalization (rather then localization), is
physically related to destructive interference of time reversed paths
for particles with spin $s=1/2$.

\subsubsection{Chiral classes: Vanishing $\beta$-function}
\label{s6.1.2}

Another peculiarity of the perturbative $\beta$-function takes place
for three chiral classes -- AIII, BDI, ad CII.  Specifically, for
these classes $\beta(t)\equiv 0$ to all orders of the perturbation
theory, as was first discovered by Gade and Wegner
\cite{gade91,gade93}.
As a result, the conductance is not renormalized at all,
serving as an exactly marginal coupling.  There is thus a line of
critical points for these models, labeled by the value of the
conductance. In fact, the $\sigma$-models for these classes contain an
additional term \cite{gade91,gade93} that does not affect the
absence of renormalization of the conductance but is crucial for the analysis
of the behavior of the DOS.

\subsubsection{Broken time-reversal invariance: Topological
$\theta$-term and quantum Hall criticality}
\label{s6.1.3}

For several classes, the $\sigma$-model action allows for inclusion of
a topological term, which is invisible to any order of the
perturbation theory. This is the case when the second homotopy group
$\pi_2$  of the $\sigma$-model manifold ${\cal M}$ (a group of homotopy
classes of maps of the sphere $S^2$ into ${\cal M}$) is
non-trivial.
From this point of view, only the compact sector ${\cal M}_F$
(originating from the fermionic part of the supervector field) of the
manifold base matters. There are five classes, for which
$\pi_2({\cal M}_F)$ is non-trivial, namely A, C, D, AII, and CII.

For the classes A, C, D the homotopy group $\pi_2({\cal M}_F) = \mathbb{Z}$.
Therefore, the action $S[Q]$ may include the (imaginary) $\theta$-term,
\be
\label{e6.1}
iS_{\rm top}[Q] = i \theta N[Q]\ ,
\ee
where an integer $N[Q]$ is the winding number of the
field configuration $Q({\bf r})$. Without loss of generality, $\theta$
can be restricted to the interval $[0,2\pi]$, since the theory is
periodic in $\theta$ with the period $2\pi$.

The topological term (\ref{e6.1}) breaks the time reversal invariance,
so it may only arise in the corresponding symmetry classes.
The by far most famous case is the Wigner-Dyson unitary class (A).
As was first understood by Pruisken\cite{pruisken84},
the $\sigma$-model of this class with the topological term
(\ref{e6.1}) describes the integer quantum Hall effect (IQHE),
with the critical
point of the plateau transition corresponding to $\theta{=}\pi$.
More recently, it was understood that  counterparts of the IQHE
exist also in the Bogoliubov-de Gennes classes with broken
time-reversal invariance -- classes
C\cite{kagalovsky99,gruzberg99,senthil99,beamond02,mirlin02}
and D.\cite{cho97,senthil00,bocquet00,chalker00,mildenberger07b}
They were called {\it spin} and
{\it thermal} quantum Hall effects (SQHE and TQHE), respectively.

\subsubsection{ $\mathbb{Z}_2$ topological term}
\label{s6.1.4}

For two classes, AII and CII, the second homotopy group is
$\pi_2({\cal M}_F) = \mathbb{Z}_2$. This allows for the $\theta$-term
but $\theta$ can only take the values $0$ and $\pi$.
It has been recently shown \cite{ostrovsky07a} that the $\sigma$-model
of the Wigner-Dyson symplectic class (AII) with a $\theta{=}\pi$
topological angle arises from a model of Dirac fermions with random
scalar potential, which describes, in particular, graphene with
long-range disorder. Like in the case of quantum-Hall systems,
this topological term inhibits localization.

\subsubsection{Wess-Zumino term}
\label{s6.1.5}

Finally, one more mechanism of emergence of criticality is the
Wess-Zumino (WZ) term that may appear in
$\sigma$-models of the classes AIII, CI, and DIII. For these classes,
the compact component ${\cal M}_F$ of the manifold is the group
$H\times H/ H = H$, where $H$ is $\text{U}(n)$, $\text{Sp}(2n)$, and
$\text{O}(2n)$,
respectively. The corresponding theories are called ``principal chiral
models''. The WZ term has the following form:
\be
\label{e6.2}
iS_{\rm WZ}(g) = \frac{ik}{24\pi} \int d^2r \int_0^1 ds \:
\epsilon_{\mu\nu\lambda}
{\rm Str} (g^{-1}\partial_\mu g)
(g^{-1}\partial_\nu g)  (g^{-1}\partial_\lambda g),
\ee
where $k$ is an integer called the level of the WZW model.
The definition (\ref{e6.2}) of the WZ term requires an extension of
the $\sigma$-model field $g({\bf r}) \equiv g(x,y)$ to the third
dimension, $0\le s \le 1$, such that $g({\bf r}, 0) = 1$ and
$g({\bf r},1) = g({\bf r})$. Such an extension is always possible,
since the second homotopy group is trivial, $\pi_2(H)=0$, for all
the three classes. Further, the value of the WZ term does not depend
on the particular way the extension to the third dimension is
performed. (This becomes explicit when one calculates the variation of
the WZ term: it is expressed in terms of   $g({\bf r})$ only.) More
precisely, there is the following topological ambiguity in the
definition of $S_{\rm WZ}(g)$. Since the third homotopy group is
non-trivial,  $\pi_3(H)=\mathbb{Z}$,  $S_{\rm WZ}(g)$ is defined up
to an arbitrary additive integer $n$ times $2\pi k$.
This, however, does not affect any
observables, since simply adds the phase $nk \times 2\pi i$ to the
action.

The WZ term arises when one bosonizes certain models of Dirac fermions
\cite{witten84} and is a manifestation of the chiral anomaly.
In particular, a $\sigma$-model for a system of
the AIII (chiral unitary) class  with the WZ term describes
Dirac fermions in a random vector potential. In this case
the $\sigma$-model coupling constant is truly marginal (as is
typical for chiral classes) and one finds a line of fixed
points. On the other hand, for the class CI there is a single fixed
point. The WZW models of these classes were encountered in the
course of study of dirty $d$-wave superconductors
\cite{nersesyan95,altland02} and, most recently, in the
context of disordered graphene. We will discuss
critical properties of these models in Sec.~\ref{s6.7.3}.

\subsection{Disordered Dirac Hamiltonians and graphene}
\label{s6.7}

Localization and criticality in models of 2D Dirac fermions subjected
to various types of disorder have been studied in a large number of
papers and in a variety of contexts, including the
random bond Ising model \cite{dotsenko83a}, the quantum Hall effect
\cite{ludwig94}, dirty superconductors with unconventional pairing
\cite{nersesyan95,bocquet00,altland02}, and some lattice models with
chiral symmetry \cite{guruswamy99}. Recently, this class of problems
has attracted a great deal of
attention\cite{mccann06,khveshchenko06,aleiner06,altland06,ostrovsky06,ostrovsky07a,ostrovsky07b}
in connection with its
application to graphene.\cite{Geim07,RMP07}

One of the most prominent
experimentally discovered features of
graphene is the ``minimal conductivity'' at the neutrality (Dirac) point.
Specifically, the conductivity\cite{Novoselov05, Zhang05,Kim} of an
undoped sample
is close to $e^2/h$ per spin per valley, remaining almost constant in
a very broad temperature range---from room temperature down to 30mK.
This is in contrast with conventional 2D systems driven by Anderson
localization into insulating state at low $T$ and suggests that
delocalization (and, possibly, quantum criticality) may emerge in a
broad temperature range due to special character of disordered
graphene Hamiltonian.

In the presence of different types of randomness, Dirac Hamiltonians
realize all ten symmetry classes of disordered systems; see
Ref. \refcite{bernard01b} for a detailed symmetry classification. Furthermore,
in many cases the Dirac character of fermions induces non-trivial
topological properties ($\theta$-term or WZ term) of the
corresponding field theory ($\sigma$-model). In Sec.~\ref{s6.7.1} we
review
the classification of disorder in  a two-flavor model of Dirac
fermions describing the low-energy physics of graphene and types of criticality.
The emergent critical theories will be discussed in
Sec.~\ref{s6.7.2}--\ref{s6.7.4}.

\subsubsection{Symmetries of  disorder and types of criticality.}
\label{s6.7.1}

The presentation below largely follows
Refs. \refcite{ostrovsky06,ostrovsky07b}. We concentrate on a two-flavor
model, which is in particular relevant to the description of
electronic properties of graphene. Graphene is a semimetal;
its valence and conduction bands touch each other in two conical
points $K$ and $K'$ of the Brillouin zone. In the
vicinity of these points the electrons behave as massless relativistic
(Dirac-like) particles. Therefore,
the effective tight-binding low-energy Hamiltonian
of clean graphene is a $4 \times 4$ matrix
operating in the $AB$ space of the two sublattices and in the
$K$--$K'$ space of
the valleys:
\begin{equation}
 H = v_0 \tau_3 \bm{\sigma}\mathbf{k}.
 \label{ham}
\end{equation}
Here $\tau_3$ is the third Pauli matrix in the $K$--$K'$ space, $\bm{\sigma}
= \{\sigma_1, \sigma_2\}$ the two-dimensional vector of Pauli matrices in
the $AB$ space, and $v_0$ the velocity  ($v_0 \simeq 10^8$ cm/s in graphene).
It is worth emphasizing that the Dirac form of the Hamiltonian (\ref{ham}) does
 not rely on the tight-binding approximation but is protected by the
 symmetry of the honeycomb lattice which has two atoms in a unit cell.

Let us analyze the symmetries of the clean
Hamiltonian (\ref{ham}) in the $AB$ and $KK'$ spaces. First,
there exists an SU(2) symmetry group in the
space of the valleys, with the generators \cite{mccann06}
\begin{equation}
 \Lambda_x
  = \sigma_3 \tau_1,
 \qquad
 \Lambda_y
  = \sigma_3 \tau_2,
 \qquad
 \Lambda_z
  = \sigma_0 \tau_3\ ,
 \label{Lambda-matrices}
\end{equation}
all of which commute with the Hamiltonian.
Second, there are two more symmetries of the clean Hamiltonian,
namely,
time inversion operation ($T_0$) and chiral symmetry ($C_0$).
Combining $T_0$, $C_0$, and isospin rotations $\Lambda_{0,x,y,z}$,
one can construct twelve symmetry operations, out of which four
(denoted as $T_\mu$) are of
time-reversal type, four ($C_\mu$) of chiral type, and four ($CT_\mu$)
of Bogoliubov-de Gennes type:
\begin{eqnarray*}
 && T_0: \ A \mapsto \sigma_1 \tau_1 A^T \sigma_1 \tau_1, \ \ \
 C_0: \ A \mapsto -\sigma_3 \tau_0 A \sigma_3 \tau_0,
 \ \ \  CT_0:\  A
  \mapsto -\sigma_2 \tau_1 A^T \sigma_2 \tau_1, \\
 && T_x:\  A \mapsto \sigma_2 \tau_0 A^T \sigma_2 \tau_0, \ \ \
 C_x:\ A \mapsto -\sigma_0 \tau_1 A \sigma_0 \tau_1,
 \ \ \  CT_x:\  A \mapsto -\sigma_1 \tau_0 A^T \sigma_1 \tau_0, \\
&&  T_y:\  A \mapsto \sigma_2 \tau_3 A^T \sigma_2 \tau_3, \ \ \
 C_y:\  A \mapsto -\sigma_0 \tau_2 A \sigma_0 \tau_2,
\ \ \  CT_y:\  A \mapsto -\sigma_1 \tau_3 A^T \sigma_1 \tau_3, \\
&& T_z:\  A \mapsto \sigma_1 \tau_2 A^T \sigma_1 \tau_2, \ \ \
 C_z:\ A \mapsto -\sigma_3 \tau_3 A \sigma_3 \tau_3,
\ \ \  CT_z:\  A \mapsto -\sigma_2 \tau_2 A^T \sigma_2 \tau_2.
\end{eqnarray*}
It is worth recalling that the $C$ and $CT$ symmetries apply to the
Dirac point
($E=0$), i.e. to undoped graphene, and get broken by a
non-zero energy $E$. We
will assume the average isotropy of the disordered graphene,
which implies that $\Lambda_x$ and $\Lambda_y$
symmetries of the Hamiltonian are present or absent
simultaneously. They are thus
combined into a single notation $\Lambda_\perp$; the same applies to
$T_\perp$ and $C_\perp$. In Table \ref{Tab:sym} all possible
matrix structures of disorder along with their symmetries are listed.

\begin{table*}
\tbl{Disorder symmetries in graphene.
The first five rows represent disorders preserving the time reversal symmetry
$T_0$;  the last four ---  violating $T_0$.
First column: structure of disorder in the sublattice ($\sigma_\mu$)
and valley ($\tau_\nu$) spaces. 
The remaining columns indicate which symmetries of the clean
Hamiltonian are preserved by disorder.  \cite{ostrovsky06}.}
{\begin{tabular}{cc@{\,}cc@{\,}c@{\,}cc@{\,}c@{\,}cc@{\,}c@{\,}c}
\hline\hline
 structure\  & 
 \makebox[0.7cm]{$\Lambda_\perp$} & \makebox[0.7cm]{$\Lambda_z$} &
 \makebox[0.7cm]{$T_0$} & \makebox[0.7cm]{$T_\perp$} & \makebox[0.7cm]{$T_z$} &
 \makebox[0.7cm]{$C_0$} & \makebox[0.7cm]{$C_\perp$} & \makebox[0.7cm]{$C_z$} &
 \makebox[0.7cm]{$CT_0$} & \makebox[0.7cm]{$CT_\perp$} & \makebox[0.7cm]{$CT_z$}
\\ \hline
 $\sigma_0 \tau_0$ & 
 $+$ & $+$ &
 $+$ & $+$ & $+$ &
 $-$ & $-$ & $-$ &
 $-$ & $-$ & $-$
\\
 $\sigma_{\{1,2\}} \tau_{\{1,2\}}$ & 
 $-$ & $-$ &
 $+$ & $-$ & $-$ &
 $+$ & $-$ & $-$ &
 $+$ & $-$ & $-$
\\
 $\sigma_{1,2} \tau_0$ & 
 $-$ & $+$ &
 $+$ & $-$ & $+$ &
 $+$ & $-$ & $+$ &
 $+$ & $-$ & $+$
\\
 $\sigma_0 \tau_{1,2}$ & 
 $-$ & $-$ &
 $+$ & $-$ & $-$ &
 $-$ & $-$ & $+$ &
 $-$ & $-$ & $+$
\\
 $\sigma_3 \tau_3$ & 
 $-$ & $+$ &
 $+$ & $-$ & $+$ &
 $-$ & $+$ & $-$ &
 $-$ & $+$ & $-$
\\ \hline
 $\sigma_3 \tau_{1,2}$ & 
 $-$ & $-$ &
 $-$ & $-$ & $+$ &
 $-$ & $-$ & $+$ &
 $+$ & $-$ & $-$
\\
 $\sigma_0 \tau_3$ & 
 $-$ & $+$ &
 $-$ & $+$ & $-$ &
 $-$ & $+$ & $-$ &
 $+$ & $-$ & $+$
\\
 $\sigma_{1,2} \tau_3$ & 
 $+$ & $+$ &
 $-$ & $-$ & $-$ &
 $+$ & $+$ & $+$ &
 $-$ & $-$ & $-$
\\
 $\sigma_3 \tau_0$ & 
 $+$ & $+$ &
 $-$ & $-$ & $-$ &
 $-$ & $-$ & $-$ &
 $+$ & $+$ & $+$
\\ \hline\hline
\end{tabular}}
\label{Tab:sym}
\end{table*}


If all types of disorder are present (i.e. no symmetries is
preserved), the RG flow is towards the
conventional localization fixed point (unitary Wigner-Dyson class A).
If the only preserved symmetry is the time reversal ($T_0$), again the
conventional localization (orthogonal Wigner-Dyson class AI) takes
place \cite{aleiner06}.  A non-trivial situation occurs if either (i) one
of the chiral symmetries is preserved or (ii) the valleys remain
decoupled. In Table \ref{Tab:result} we list situations when symmetry
prevents localization and leads to criticality and non-zero conductivity at
$E{=}0$ (in the case of decoupled nodes  -- also at nonzero
$E$). Models with decoupled nodes are analyzed in
Sec.~\ref{s6.7.2}, and models with a chiral symmetry in
Sec.~\ref{s6.7.3} ($C_0$-chirality) and \ref{s6.7.4}
($C_z$-chirality).

\begin{table*}
\tbl{Possible types of disorder in graphene leading to
  criticality.
The first three row correspond to $C_z$ chiral symmetry leading to
Gade-Wegner-type criticality, Sec.~\protect\ref{s6.7.4}. The next
three rows contain models with $C_0$ chiral symmetry (random gauge
fields), inducing a WZ term in the $\sigma$-model action,
Sec.~\protect\ref{s6.7.3}. The last four rows correspond to the
case of decoupled valleys (long-range disorder), see
Sec.~\protect\ref{s6.7.2}; In the last three cases the
$\sigma$-model acquires a topological term with $\theta=\pi$.
Adapted from Ref. \refcite{ostrovsky07b}. }
{\begin{tabular*}{30.2pc}{ccccc}
\hline\hline
Disorder &
\hspace*{-0.5cm} Symmetries \hspace*{-0.5cm}             & Class      
    & \hspace*{-0.5cm} Criticality\hspace*{-0.5cm}  &
Conductivity \\
\hline
\hspace*{-0.4cm}Vacancies, strong potential impurities &
$C_z$, $T_0$            & BDI            & Gade &
$\approx 4e^2/\pi h$ \\
Vacancies + RMF &
$C_z$                   & AIII           & Gade &
$\approx 4e^2/\pi h$ \\
$\sigma_3\tau_{1,2}$ disorder &
$C_z$, $T_z$            & CII            & Gade &
$\approx 4e^2/\pi h$ \\
\hline
Dislocations &
$C_0$, $T_0$            & CI             & WZW &
$4e^2/\pi h$ \\
Dislocations + RMF &
$C_0$                   & AIII           & WZW &
$4e^2/\pi h$ \\
\hline
Ripples, RMF &
$\Lambda_z$, $C_0$      & $2 \times$AIII & WZW &
$4e^2/\pi h$ \\
\hline
Charged impurities &
$\Lambda_z$, $T_\perp$  & $2 \times$AII  & $\theta = \pi$ &
$4\sigma_{Sp}^{**}$ or$^{\text a}$ $(4e^2/\pi h)\ln L$ \\
Random Dirac mass: $\sigma_3\tau_0$, $\sigma_3\tau_3$ &
$\Lambda_z$, $CT_\perp$ & $2 \times$D    & $\theta = \pi$ &
$4e^2/\pi h$ \\
Charged impurities + (RMF, ripples) &
$\Lambda_z$             & $2 \times$A    & $\theta = \pi$ &
$4\sigma_U^*$ \\
\hline\hline
\end{tabular*}}
\begin{tabnote}
$^{\text a}$Numerical simulations\cite{Bardarson07} reveal a flow towards the
supermetal fixed point, $\sigma\simeq(4e^2/\pi h)\ln L \to \infty$.  
\end{tabnote}
\label{Tab:result}
\end{table*}

\subsubsection{Decoupled nodes: Disordered single-flavor Dirac
  fermions and quantum-Hall-type criticality}
\label{s6.7.2}

If the disorder is of long-range character,  the valley
mixing is absent due to the lack of scattering with large momentum
transfer.
For each of the nodes, the system can then be described
in terms of a single-flavor Dirac Hamiltonian,
\begin{equation}
 H
  = v_0 [\bm{\sigma}\mathbf{k} + \sigma_\mu V_\mu(\mathbf{r})].
 \label{ham1}
\end{equation}
Here disorder includes random scalar ($V_0$) and vector ($V_{1,2}$) potentials
and random mass ($V_3$). 
The clean single-valley Hamiltonian (\ref{ham1}) obeys
the
effective time-reversal invariance $H = \sigma_2 H^T\sigma_2$. This symmetry
($T_\perp$) is not the physical time-reversal symmetry ($T_0$): the latter
interchanges the nodes and is of no significance in the absence of inter-node
scattering.

Remarkably, single-flavor Dirac fermions are never in the conventional
localized phase! More specifically,
depending on which of the disorders are present, four different types
of criticality take place:

(i) The only disorder is the random vector potential ($V_{1,2}$). This is
a special case of the symmetry class AIII. This problem is exactly
solvable. It is characterized by a line of fixed points, all showing
conductivity $4e^2/\pi h$, see Sec.~\ref{s6.7.3}.

(ii) Only random mass ($V_3$) is present. The system belongs then to
class D. The random-mass disorder is marginally irrelevant, and the
system flows under RG towards the clean fixed point, with the
conductivity $4e^2/\pi h$.

(iii) The only disorder is random scalar potential ($V_0$). The system
is then in the Wigner-Dyson symplectic (AII) symmetry class. As was
found in Ref.~\refcite{ostrovsky07a}, the corresponding $\sigma$-model
contains a  $\mathbb{Z}_2$ topological term with $\theta=\pi$ which
protects the system from localization. The absence of
localization in this model has been confirmed in numerical
simulations.\cite{Bardarson07}
The scaling function has been found in Ref.~\refcite{Bardarson07}
to be strictly positive, implying a flow towards the ``supermetal''
fixed point.

(iv) At least two types of randomness are present. 
All symmetries are broken in this case and the model belongs to the
Wigner-Dyson unitary class A. It was argued in
Ref. \refcite{ludwig94} that it flows into the IQH transition
fixed point. This is confirmed by the derivation of the corresponding
$\sigma$-model \cite{altland02,ostrovsky07a,ostrovsky07b}, which contains a
topological term with $\theta=\pi$, i.e. is nothing but
the Pruisken $\sigma$-model at criticality. A particular consequence
of this is that the conductivity of graphene with this type of
disorder is equal to the value $\sigma_U^*$ of the longitudinal
conductivity $\sigma_{xx}$ at the critical point of the IQH transition
multiplied by four (because of spin and valleys).

If a uniform transverse magnetic field is applied, the topological angle
$\theta$ becomes energy-dependent. However, at the Dirac point
($E = 0$), where $\sigma_{xy}
= 0$, its value remains unchanged, $\theta = \pi$. This implies the emergence
of the half-integer quantum Hall effect, with a plateau transition point at
$E = 0$.

\subsubsection{Preserved $C_0$ chirality:  Random gauge fields}
\label{s6.7.3}

Let us consider a type of disorder which preserves the $C_0$-chirality, $H =
-\sigma_3 H \sigma_3$. This implies the disorder of the type $\sigma_{1,2}
\tau_{0,1,2,3}$ being strictly off-diagonal in the $\sigma$ space.
Depending on further symmetries, three different $C_0$-chiral
models arise:

(i) The only disorder present is $\sigma_{1,2}\tau_3$, which corresponds to
the random abelian vector potential. In this case the
nodes are decoupled, and the Hamiltonian decomposes in two copies of
a model of the class AIII. This model characterized by a line of fixed
points has already been mentioned in
Sec.\ref{s6.7.2}.

(ii) If the time-reversal symmetry $T_0$ is preserved, only the
disorder of the type $\sigma_{1,2}\tau_{0,1,2}$ is allowed, and the
problem is in the symmetry class CI. The model
describes then fermions coupled to a SU(2) non-abelian gauge field,
and is a particular case of analogous SU(N) models. This
theory flows now into an isolated fixed
point, which is a WZW theory on the level
$k=-2N$.\cite{nersesyan95,mudry96,Caux}

(iii) All $C_0$-invariant disorder structures are present. 
This describes Dirac fermions
coupled to both abelian U(1) and non-abelian SU(2) gauge fields. This
model is in the AIII symmetry class.

Remarkably, all these critical $C_0$-chiral models are exactly
solvable. In particular, the critical conductivity can be calculated
exactly and is independent on the disorder strength. A general proof of this
statement based only on the gauge invariance is given in Ref.\
\refcite{ostrovsky06}.
(For particular cases it was earlier obtained in
Refs.\refcite{ludwig94,tsvelik95}). The critical conductivity is thus the same
as in clean graphene,
\begin{equation}
 \sigma
  = 4 e^2 / \pi h.
 \label{universal}
\end{equation}
Spectra of multifractal exponents and the critical index
of the DOS can also be calculated exactly, see the
review\cite{evers08}.

\subsubsection{Disorders preserving $C_z$ chirality:  Gade-Wegner
criticality}
\label{s6.7.4}

Let us now turn to the disorder which preserves the $C_z$-chirality,
$H = -\sigma_3 \tau_3 H \sigma_3 \tau_3$; according to Table~\ref{Tab:sym},
the corresponding disorder structure is $\sigma_{1,2} \tau_{0,3}$ and
$\sigma_{0,3} \tau_{1,2}$
If no time-reversal symmetries are preserved, the system
belongs to the chiral unitary (AIII) class.
The combination of $C_z$-chirality and the time reversal invariance
$T_0$ 
corresponds to the
chiral orthogonal
symmetry class BDI; this model has already been discussed in
Sec.~\ref{s6.1.2}.  Finally, the combination of $C_z$-chirality and
$T_z$-symmetry 
falls into the
chiral symplectic symmetry class CII. The RG flow and DOS in these
models have been analyzed in Ref. \refcite{guruswamy99} 
In all the cases, the
resulting theory is of the Gade-Wegner type.\cite{gade91,gade93}
These theories are characterized by lines of fixed points, with
non-universal conductivity. It was found\cite{ostrovsky06,ryu07a} that for weak
disorder the conductivity takes approximately the universal value,
$\sigma \simeq 4e^2 /\pi h$. In contrast to the case of $C_0$
chirality, this result is, however, not exact. In particular, the leading
correction to the clean conductivity is found in the second order
in disorder strength\cite{ostrovsky06}.

\section{Electron-electron-interaction effects}
\label{s7.1}

Physically, the impact of interaction effects onto low-temperature
transport and localization in
disordered electronic systems can be subdivided into two distinct
effects: (i) renormalization and (ii) dephasing.

\paragraph{Renormalization.}
The renormalization
effects, which are governed by virtual processes, become increasingly
more pronounced with lowering temperature. The importance of such
effects in diffusive low-dimensional systems was demonstrated
by Altshuler and Aronov, see Ref.~\refcite{altshuler84}. To resum
the arising singular contributions, Finkelstein developed the RG
approach based on the $\sigma$-model for an interacting system, see
Ref.~\refcite{finkelstein90} for a review. This made possible an analysis of the
critical behavior at the localization transition in $2+\epsilon$
dimensions in the situations when spin-rotation invariance is broken (by
spin-orbit scattering, magnetic field, or magnetic impurities).
However, in the case of preserved spin-rotation symmetry
it was found that the strength of the interaction
in spin-triplet channel scales to infinity at certain RG scale.
This was interpreted as some kind of magnetic instability of the
system; for a detailed exposition of proposed scenarios
see Ref. \refcite{belitz94}.

Recently, the problem has attracted a
great deal of attention in connection with experiments on
high-mobility low-density 2D electron structures (Si MOSFETs) giving an
evidence in favor of a metal-insulator transition \cite{abrahams01}.
In Ref.~\refcite{punnoose05} the RG for $\sigma$-model
for interacting 2D electrons with a number of valleys $N>1$ was
analyzed on the two-loop level. It was shown that in
the limit of large number of valleys $N$ (in practice, $N=2$ as in Si
is already sufficient) the temperature of magnetic instability is
suppressed down to unrealistically low temperatures, and a
metal-insulator transition emerges. The existence of
interaction-induced metallic phase in 2D is due to the fact that, for
a sufficiently strong interaction, its ``delocalizing'' effect
overcomes the disorder-induced localization. Recent
works~\cite{knyazev08,punnoose09} show that the RG theory describes well
the experimental data up to lowest accessible temperatures.
We will see in Sec.~\ref{s8} that the Coulomb interaction may also
lead to dramatic effects in the context of topological insulators.

The interaction-induced renormalization effects become extremely
strong for correlated 1D systems (Luttinger liquids). While 1D systems
provide a paradigmatic example of strong Anderson localization,  a sufficiently
strong attractive interaction can lead to delocalization in such
systems. An RG treatment of the corresponding localization transition in
a disordered interacting 1D systems was developed in
Ref.~\refcite{giamarchi88}, see also the book \refcite{giamarchi04}.
Recently, the interplay between Anderson localization, Luttinger-liquid
renormalization, and dephasing has been studied in detail in Ref.~\refcite{gornyi07}.

\paragraph{Dephasing.}
We turn now to effects of dephasing governed by inelastic processes of
electron-electron scattering at
finite temperature $T$.  The dephasing has been studied in great
detail for metallic systems where it provides a cutoff for weak
localization effects \cite{altshuler84}. As to the Anderson
transitions, they are quantum (zero-$T$) phase transitions, and dephasing
contributes to their smearing at finite $T$. The dephasing-induced
width of the transition scales as a power-law function of $T$.
There is, however, an interesting situation when
dephasing processes can create a localization transition. We mean
the systems where all states are localized in the absence of
interaction, such as wires or 2D systems. At high temperatures,
when the dephasing is strong, so that the dephasing rate
$\tau_\phi^{-1}(T)$ is larger than mean level spacing in the
localization volume, the system is a good metal and its conductivity
is given by the quasiclassical Drude conductivity with relatively
small weak localization correction \cite{altshuler84}. With lowering
temperature the dephasing gets progressively less efficient, the
localization effects proliferate, and eventually the system becomes an
Anderson insulator. What is the nature of this state? A natural
question is whether the interaction of an electron with other
electrons will be
sufficient to provide a kind of thermal bath that would assist the
variable-range hopping transport \cite{fleishman78},
as it happens in the presence of a phonon bath. The answer to this
question was given by Ref.~\refcite{fleishman80}, and
it is negative. Fleishman and Anderson found that at low $T$ the
interaction of a ``short-range class'' (which includes a finite-range
interaction in any dimensionality $d$ and Coulomb interaction in
$d<3$) is not sufficient to delocalize otherwise localized electrons,
so that the conductivity remains strictly zero. In combination with
the Drude conductivity at high-$T$ this implies the existence of
transition at some temperature $T_c$.

This conclusion was recently
corroborated by an analysis \cite{gornyi05,basko06} in the framework
of the idea of Anderson localization in Fock space \cite{altshuler97a}.
In these works the temperature dependence of conductivity $\sigma(T)$
in systems with localized states and weak electron-electron
interaction was studied. It was found that with decreasing $T$ the
system first shows a crossover from the weak-localization regime
into that of ``power-law hopping'' over localized states
(where $\sigma$ is a power-law function of $T$), and then undergoes a
localization transition. The transition  is obtained both within a
self-consistent Born approximation \cite{basko06} and an approximate
mapping onto a model on the
Bethe lattice \cite{gornyi05}. The latter yields also
a critical behavior of $\sigma(T)$ above $T_c$, which has a
characteristic for the Bethe lattice non-power-law form $\ln\sigma(T)
\sim (T-T_c)^{-\kappa}$ with $\kappa=1/2$, see Sec.~\ref{s2.4}.

Up to now, this
transition has not been observed in experiments\footnote{Of course,
in a real system, phonons are always present and provide a bath
necessary to support the hopping conductivity at low $T$, so that
there is no true transition. However, when the coupling to phonons is
weak, this hopping conductivity will have a small prefactor, yielding
a ``quasi-transition''.}, which indicate
instead a smooth crossover from the metallic to the insulating phase
with lowering $T$ \cite{hsu95,vankeuls97,khavin98,minkov07}.
The reason for this discrepancy remains unclear.
An attempt to detect the transition in numerical simulations
also did not give a clear confirmation of the
theory~\cite{oganesyan07}, possibly because of
strong restrictions on the size of an interacting system that can be
numerically diagonalized. On the other hand, a very recent work
\cite{Monthus10} does report an evidence in favor of a transition of a
Bethe-lattice character (though with different value of $\kappa$).

\section{Topological Insulators}
\label{s8}

One of the most recent arenas where
novel peculiar localization phenomena have been studied is physics of
topological 
insulators~\cite{Kane05, Bernevig06, Zhang08, Koenig07,
Hsieh08, Roth09, Hsieh09, Hsieh09a}.
Topological insulators are bulk insulators with delocalized
(topologically protected) 
states on their surface.\footnote{Related topology-induced phenomena
have been considered in Ref.~\refcite{Volovik} in the context of superfluid 
Helium-3 films). }
As discussed above, the critical behavior of a system depends on
the underlying topology.  This is particularly relevant for
topological insulators.

The famous example of a topological insulator is a two-dimensional
(2D) system on one of quantum Hall plateaus in the integer quantum
Hall effect. Such a system is characterized by an integer (Chern
number) $n = ...,-2,-1,0,1,2,...$ which counts the edge states
(here the sign determines the direction of chiral edge modes). The
integer quantum Hall edge is thus a topologically protected
one-dimensional (1D) conductor realizing the group $\mathbb{Z}$.

Another ($\mathbb{Z}_2$) class of topological insulators
\cite{Kane05, Bernevig06, Zhang08} can be realized in systems with
strong  spin-orbit interaction and without magnetic field (class AII)
--- and was discovered in 2D HgTe/HgCdTe structures in
Ref.~\refcite{Koenig07} (see also Ref.~\refcite{Roth09}). A 3D
$\mathbb{Z}_2$ topological insulator \cite{Hsieh08} has been found
and investigated for the first time in Bi$_{1-x}$Sb$_{x}$
crystals. Both in 2D and 3D, $\mathbb{Z}_2$ topological insulators
are band insulators with the following properties: (i) time
reversal invariance is preserved (unlike ordinary quantum Hall
systems); (ii) there exists a topological invariant, which is
similar to the Chern number in QHE; (iii) this invariant belongs
to the group $\mathbb{Z}_2$ and reflects the presence or absence
of delocalized edge modes (Kramers pairs)~\cite{Kane05}.

Topological insulators 
exist in all ten symmetry classes in different dimensions, see Table
\ref{Tab:TI}. Very generally, the
classification of topological insulators in $d$ dimensions can be
constructed by studying the Anderson localization problem in a
$(d-1)$-dimensional disordered system \cite{Schnyder08}. Indeed, absence of
localization of surface states due to the topological protection
implies the topological character of the insulator.

In Sec.~\ref{s8.1} we overview the full classification of topological
insulators and superconductors~\cite{Schnyder08,Kitaev09}.
In Sec.~\ref{s8.2} we discuss $\mathbb{Z}_2$ topological insulators
belonging to the symplectic  
symmetry class AII, characteristic to systems with strong spin-orbit
interaction. Finally, in Sec.~\ref{s8.3} we address, closely following
Ref.~\refcite{OGM-TI09}, 
the interaction effects
in $\mathbb{Z}_2$ topological insulators.

\subsection{Symmetry classification of topological insulators}
\label{s8.1}

\begin{table*}
\tbl{Symmetry classes and ``Periodic Table'' of topological
insulators \cite{Kitaev09,Schnyder08}. The first column enumerates
the symmetry classes of disordered systems which are defined as
the symmetry classes $H_p$ of the Hamiltonians (second column).
The third column lists the symmetry classes of the classifying
spaces (spaces of reduced Hamiltonians) \cite{Kitaev09}. The
fourth column represents the symmetry classes of a compact sector
of the sigma-model manifold. The fifth column displays the zeroth
homotopy group $\pi_0(R_p)$ of the classifying space. The last
four columns show the possibility of existence of $\mathbb{Z}$ and
$\mathbb{Z}_2$ topological insulators in each symmetry class in
dimensions $d=1,2,3,4$. Adapted from Ref. \refcite{OGM-TI09}.}
{\begin{tabular}{clclclclclclclclc}
\hline\hline
 \multicolumn{1}{c}{} &&
 \multicolumn{3}{c}{Symmetry classes} &&
\multicolumn{1}{c}{} &&
 \multicolumn{4}{c}{Topological insulators}
\\
 $p$ &&
 $H_p$ & $R_p$ & $S_p$ &&
 $\pi_0(R_p)$ &&
 d=1 & d=2 & d=3 & d=4
\\ \hline\hline
 0
 &&
 AI & BDI & CII
 &&
$\mathbb{Z}$
 &&
 0 & 0 & 0 & $\mathbb{Z}$
\\
 1
&&
 BDI & BD & AII
&& $\mathbb{Z}_2$
  &&
 $\mathbb{Z}$ & 0 & 0 & 0
\\
 2
&&
 BD & DIII  & DIII
&& $\mathbb{Z}_2$
  &&
 $\mathbb{Z}_2$ & $\mathbb{Z}$ & 0 & 0
\\
3
 &&
 DIII & AII & BD &&
 0 &&
$\mathbb{Z}_2$ & $\mathbb{Z}_2$ & $\mathbb{Z}$ & 0
\\
 4
&&
 AII & CII & BDI
&&
 $\mathbb{Z}$
&& 0 & $\mathbb{Z}_2$ & $\mathbb{Z}_2$ & $\mathbb{Z}$
\\
 5
&&
 CII & C & AI
&&
 0
&& $\mathbb{Z}$ & 0 & $\mathbb{Z}_2$ & $\mathbb{Z}_2$
\\
 6
&&
 C & CI & CI
&&
 0
&& 0 & $\mathbb{Z}$ & 0 & $\mathbb{Z}_2$
\\
 7
&&
 CI & AI & C
&&
 0
&& 0 & 0 & $\mathbb{Z}$ & 0
\\
\hline  $0^\prime$ &&
 A & AIII & AIII
&&
 $\mathbb{Z}$
&& 0 & $\mathbb{Z}$ & 0 &  $\mathbb{Z}$
\\
$1^\prime$ &&
 AIII & A & A
&& 0 && $\mathbb{Z}$ & 0 & $\mathbb{Z}$ & 0\\
\hline\hline
\end{tabular}}
\label{Tab:TI}
\end{table*}

The full classification (periodic table) of topological insulators
and superconductors for all ten symmetry classes
\cite{zirnbauer96b,altland97} was developed in Refs.\
\refcite{Kitaev09} and \refcite{Schnyder08}. This classification
determines whether the $\mathbb{Z}$ or $\mathbb{Z}_2$ topological
insulator is possible in the $d$-dimensional system of a given
symmetry class. In this Section we overview the classification of
topological insulators closely following  Refs. \refcite{Kitaev09}
and \refcite{Schnyder08}, and discuss the connection between the
classification schemes of these papers. 

All symmetry classes of disordered systems
(see Section~\ref{s:SymDisSys} and
Table~\ref{t4.1}) can be divided into two groups: \{A,AIII\} and
\{all other\}. The classes of the big group are labeled by
$p=0,1,\dots,7$. Each class is characterized by (i) Hamiltonian
symmetry class $H_p$; (ii) symmetry class $R_p$ of the classifying
space used by Kitaev \cite{Kitaev09}; (iii) symmetry class $S_p$
of the compact sector $\mathcal{M}_F$ of the sigma-model manifold.
The symmetry class $R_p$  of the classifying space of reduced
Hamiltonians characterizes the space of matrices obtained from the
Hamiltonian by keeping all eigenvectors and replacing all positive
eigenvalues by $+1$ and all negative by $-1$. Note that
\begin{equation}
 R_p=H_{p+1}, \quad S_p=R_{4-p}.
\label{RHS}
\end{equation}
Here and below cyclic definition of indices $\{0,1,\dots,7\}$ (mod
8) and $\{0',1'\}$ (mod 2) is assumed.

For the classification of topological insulators it is important
to know homotopy groups $\pi_d$ for all symmetry classes. In Table
\ref{Tab:TI} we list $\pi_0(R_p)$; other $\pi_d$ are given by
\begin{equation}
 \pi_d(R_p)=\pi_0(R_{p+d}).
\label{pidRp}
\end{equation}
The homotopy groups $\pi_d$ have periodicity 8 (Bott periodicity).

There are two ways to detect topological insulators:
by inspecting the topology of (i) classifying space $R_p$
or of (ii) the sigma-model space $S_p$.
\begin{enumerate}
\item[(i)]
Existence of topological insulator (TI) of class $p$ in $d$ dimensions
is established by the homotopy group $\pi_0$ for the classifying space  $R_{p-d}$:
\begin{equation}
\begin{cases}
       \text{TI of the type}\ \mathbb{Z}   \\
        \text{TI of the type}\ \mathbb{Z}_2  \end{cases}
\Longleftrightarrow \quad
\pi_0(R_{p-d}) = \begin{cases}
        \mathbb{Z} &  \\
          \mathbb{Z}_2 &
                  \end{cases}
\end{equation}

\item[(ii)]
Alternatively, the existence of topological insulator of
symmetry class $p$ in $d$ dimensions can be inferred from the
homotopy groups of the sigma-model manifolds, as follows:
\begin{equation}
\begin{cases}
       \text{TI of the type}\ \mathbb{Z} \ \Longleftrightarrow  &\pi_d(S_{p}) = \mathbb{Z}\\
        \text{TI of the type}\ \mathbb{Z}_2  \Longleftrightarrow  &\pi_{d-1}(S_{p}) =\mathbb{Z}_2
\end{cases}
\end{equation}
\end{enumerate}

The criterion (ii) is obtained if one requires existence of
``non-localizable'' boundary excitations. This may be guaranteed
by either Wess-Zumino term in $d-1$ dimensions [which is
equivalent to the $\mathbb{Z}$ topological term in $d$ dimensions,
i.e. $\pi_d(S_{p}) = \mathbb{Z}$] for a QHE-type topological
insulator, or by the $\mathbb{Z}_2$ topological term in $d-1$
dimensions  [i.e. $\pi_{d-1}(S_{p}) = \mathbb{Z}_2$] for a
QSH-type topological insulator.

The above criteria (i) and (ii) are equivalent, since
\begin{equation}
 \pi_d(S_p)=\pi_d(R_{4-p})=\pi_0(R_{4-p+d}).
\label{pidSp}
\end{equation}
and
\begin{equation}
 \pi_0(R_p)=  \begin{cases}
        \mathbb{Z} &  \text{for}\ p=0,4,\\
          \mathbb{Z}_2 & \text{for}\ p=1,2.
                  \end{cases}
\label{pi0Rp}
\end{equation}

Below we focus on 2D systems of symplectic (AII) symmetry class.
One sees that this is the only symmetry class out of ten classes
that supports the existence of $\mathbb{Z}_2$ topological
insulators both in 2D and 3D.

\subsection{$Z_2$ topological insulators in 2D and 3D systems of class AII}
\label{s8.2}

A $\mathbb{Z}_2$ class of topological insulators belonging to the
symmetry class AII was first realized in 2D HgTe/HgCdTe structures in
Ref.~\refcite{Koenig07}. Such systems were found to possess two
distinct insulating phases, both having a gap in the bulk electron
spectrum but differing by edge properties. 
While the normal
insulating phase has no edge states, the topologically nontrivial
insulator is characterized by a pair of mutually time-reversed
delocalized edge modes penetrating the bulk gap. Such state shows
the quantum spin Hall (QSH) effect which was theoretically
predicted in a model system of graphene with spin-orbit
coupling.~\cite{Kane05, Sheng05}
The transition between the two
topologically nonequivalent phases (ordinary and QSH insulators)
is driven by inverting the band gap~\cite{Bernevig06}. The
$\mathbb{Z}_{2}$ topological order is robust with respect to
disorder: since the time-reversal invariance forbids
backscattering of edge states at the boundary of QSH insulators,
these states are topologically protected from localization.

For clean 2D QSH systems with a bulk gap generated by spin-orbit
interaction, the $\mathbb{Z}_2$ invariant can be constructed from
the Bloch wave functions on the Brillouin zone \cite{Kane05} and is
somewhat similar to the Chern number in the standard QHE. 
Formally, if the $\mathbb{Z}_{2}$ index is odd/even there is an
odd/even number $m$  
of Kramers pairs of gapless edge states (here $m=0$ is treated as even number).
In the presence of disorder which generically back-scatters between
different Kramers pairs,  
all the surface modes get localized if $m$ was even 
in the clean system, while a single delocalized pair survives if $m$ was odd.

Disorder was found to induce a metallic phase separating the two
(QSH and ordinary) insulators~\cite{Onoda07,Obuse07}. The
transition between metal and any of the two
insulators occurs at the critical value of conductivity $g=g^*
\approx 1.4$; both transitions are believed to belong to the same
universality class, see Sections \ref{s2.2.2} and \ref{s6.1.1}.
For $g<g^*$ all bulk states are eventually localized in the limit
of large system, while for $g>g^*$ the weak antilocalization
specific to the symplectic symmetry class drives the system to the
``supermetallic'' state, $g \to \infty$. The schematic phase
diagram for the noninteracting case is shown in Fig.\
\ref{Fig:ph} (left panel).

A related three-dimensional (3D) $\mathbb{Z}_2$ topological
insulator was discovered in Ref. \refcite{Hsieh08} where crystals of
Bi$_{1-x}$Sb$_{x}$ were investigated. The boundary in this case
gives rise to a 2D topologically protected metal. Similarly to 2D
topological insulators, the inversion of the 3D band gap induces
an odd number of the surface 2D modes \cite{Dyakonov81,Fu07}.
These states in BiSb have been studied experimentally in Refs.\
\refcite{Hsieh08} and \refcite{Hsieh09}. Other examples of 3D
topological insulators include BiTe and BiSe
systems~\cite{Hsieh09a}. The effective 2D surface Hamiltonian has
a Rashba form and describes a single species of 2D massless Dirac
particles (cf. Ref. \refcite{Volkov85}). It is thus analogous to the
Hamiltonian of graphene with just a single valley. In the absence
of interaction, the conductivity of the disordered surface of a 3D
topological insulator therefore scales to infinity with increasing
the system size, see Section \ref{s6.7.2}.

\subsection{Interaction effects on $Z_2$ topological insulators of class AII}
\label{s8.3}

In this Section, we overview the effect of Coulomb interaction between
electrons in topological insulators~\cite{OGM-TI09}. 
Since a topological insulator is
characterized by the presence of propagating surface modes, its
robustness with respect to interactions means that interactions do
not localize the boundary states. Indeed, arguments showing the
stability of $\mathbb{Z}_2$ topological insulators with respect to
interactions were given in Refs.~\refcite{Kane05,Zhang08,Lee08} and
\refcite{OGM-TI09}. An additional argument in favor of persistence of
topological protection in the presence of interaction is based on
the replicated Matsubara sigma-model, in analogy with the ordinary
QHE~\cite{Burmistrov}. This theory possesses the same nontrivial
topology as in the non-interacting case.

Can the topologically protected 2D state be a supermetal ($g \to
\infty$) as in the noninteracting case? To answer this question 
the perturbative RG applicable for large conductivity $g\gg
1$ has been employed in Ref.~\refcite{OGM-TI09}. 
It is well known that in a 2D diffusive system the interaction
leads to logarithmic corrections to the
conductivity~\cite{altshuler84}, see Sec.~\ref{s7.1}. These
corrections (together with the interference-induced ones) can be
summed up with the use of RG technique~\cite{finkelstein90,
belitz94}. 

The one-loop equation for renormalization of the
conductivity in the symplectic class with long-range Coulomb
interaction and a single species of particles has the following
form:
\begin{equation}
 \beta(g)=\frac{d g}{d \ln L}=-1/2.
\label{RG}
\end{equation}
Here $-1/2$ on the r.h.s. is a sum of the weak antilocalization
correction $1/2$ due to disorder and $-1$ induced by the Coulomb
interaction in the singlet channel. According to Eq.~(\ref{RG}),
the negative interaction-induced term in $\beta(g)$  dominates the
scaling at large $g$. Therefore, for $g\gg 1$ the conductance
decreases upon renormalization and the supermetal fixed point
becomes repulsive.

Thus, on one hand, at $g\gg 1$  there is (i) scaling towards
smaller $g$ on the side of large $g$. On the other hand, 
surface states are topologically protected from localization, which
yields (ii) scaling towards higher $g$ on the side of small $g$. 
The combination of
(i) and (ii) leads unavoidably to the conclusion that the system
should scale to a critical state ($g \sim 1$). Indeed, there is no other way to
continuously interpolate between negative (i) and positive (ii)
beta functions: at some point $\beta(g)$
should cross zero. As a result, a critical point emerges due to
the combined effect of interaction and topology \cite{OGM-TI09}.
In other words, if the system can flow neither towards a
supermetal ($g\to\infty$) nor to an insulator ($g\to 0$) it must
flow to an intermediate fixed point ($g\sim 1$). Remarkably, the
critical state emerges on the surface of a 3D topological
insulator without any adjustable parameters. This phenomenon can
be thus called ``self-organized quantum criticality''
\cite{OGM-TI09}.

\begin{figure}
\includegraphics[width=\columnwidth]{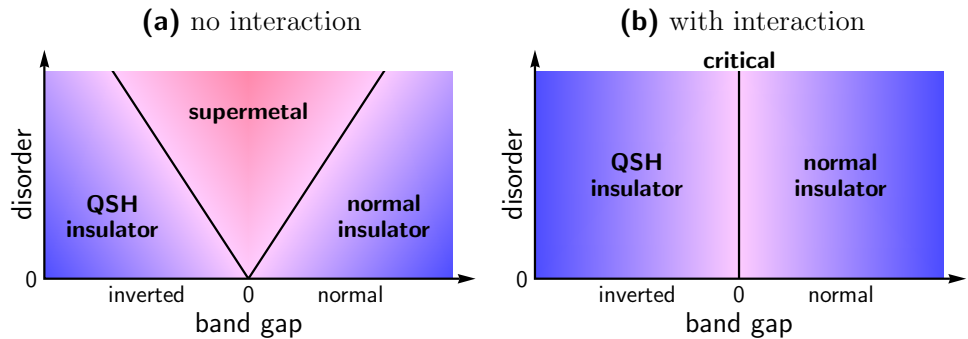}
\caption{The phase diagrams of a disordered 2D system
demonstrating the QSH effect. Left: noninteracting case.  Right:
interacting case (with Coulomb interaction not screened by
external gates). Interaction ``kills'' the supermetallic phase. As
a result, the two insulating phases are separated by the critical
line. Adapted from Ref. \refcite{OGM-TI09}.} \vspace*{-0.4cm}
\label{Fig:ph}
\end{figure}

Let us now return to 2D  $\mathbb{Z}_2$ topological insulators.
The 2D disordered QSH system contains only a single flavor of
particles, $N=1$. Indeed, the spin-orbit interaction breaks the
spin-rotational symmetry, whereas the valleys are mixed by
disorder. As a result, the supermetal phase does not survive in
the presence of Coulomb interaction: at $g\gg 1$ the
interaction-induced localization wins. This is analogous to the
case of the surface of a 3D topological insulator discussed above.

The edge of a 2D topological insulator is protected from the full
localization~\cite{Kane05}. This means that the topological
distinction between the two insulating phases (ordinary and QSH
insulator) is not destroyed by the interaction, whereas the
supermetallic phase separating them disappears. Therefore, the
transition between two insulators occurs through an
interaction-induced critical state\cite{OGM-TI09}, see Fig.\ \ref{Fig:ph} (right panel).

\section{Summary}
\label{s9}

Despite its half-a-century age, Anderson localization remains a very
actively developing field. In this article, we have reviewed some of
recent theoretical advances in the physics of Anderson transitions,
with an emphasis on
manifestations of criticality and on the impact of underlying symmetries and
topologies. The ongoing progress in experimental techniques allows one
to explore these concepts in a variety of materials, including
semiconductor structures, disordered superconductors,
graphene, topological insulators, atomic
systems, light and sound propagating in random media, etc.

We are very grateful to a great many of colleagues for fruitful
collaboration and
stimulating discussions over the years of research work in this
remarkable field.
The work was supported by the DFG -- Center for
Functional Nanostructures, by the EUROHORCS/ESF (IVG), and by Rosnauka
grant 02.740.11.5072.


\printindex
\end{document}